\documentclass[aps,prl,superscriptaddress,twocolumn]{revtex4-2}
\usepackage{eurosym}
\usepackage{amsfonts}
\usepackage{amsmath}
\usepackage{mathrsfs}
\usepackage{amssymb}
\usepackage{graphicx}
\usepackage{float}
\usepackage{comment}
\usepackage{bm}
\usepackage{color}
\usepackage[colorlinks=true,linkcolor=blue,anchorcolor=blue,citecolor=blue,urlcolor=blue]{hyperref}
\usepackage[normalem]{ulem}
\usepackage{siunitx}
\usepackage[tracking]{microtype}
\usepackage[version=4]{mhchem}
\usepackage{elements}

\SetTracking{ encoding = *, shape = sc }{ 25 }

\DeclareMathSymbol{\shortminus}{\mathbin}{AMSa}{"39}

\begin{document}

\title{Scaling of quantum Fisher information for quantum exceptional point
sensors}
\author{Chun-Hui Liu}
\affiliation{Department of Physics, Washington University, St. Louis, MO 63130, USA}
\affiliation{Department of Physics, The University of Texas at Dallas, Richardson, Texas
75080, USA}
\author{Fu Li}
\affiliation{Department of Electrical and Systems Engineering, Washington University, St.
Louis, MO 63130, USA}
\author{Shengwang Du}
\affiliation{Department of Physics, The University of Texas at Dallas, Richardson, Texas
75080, USA}
\affiliation{Elmore Family School of Electrical and Computer Engineering, Purdue University, West Lafayette, IN 47907}
\affiliation{Department of Physics and Astronomy, Purdue University, West Lafayette, IN 47907}
\author{Jianming Wen}
\affiliation{Department of Physics, Kennesaw State University, Marietta, GA 30060, USA}
\author{Lan Yang}
\affiliation{Department of Electrical and Systems Engineering, Washington University, St.
Louis, MO 63130, USA}
\author{Chuanwei Zhang}
\email{chuanwei.zhang@wustl.edu}
\affiliation{Department of Physics, Washington University, St. Louis, MO 63130, USA}
\affiliation{Department of Physics, The University of Texas at Dallas, Richardson, Texas
75080, USA}

\begin{abstract}
In recent years, significant progress has been made in utilizing the
divergence of spectrum response rate at the exceptional point (EP) for
sensing in classical systems, while the use and characterization of quantum
EPs for sensing have been largely unexplored. For a quantum EP sensor, an
important issue is the relation between the order of the quantum EP and
the scaling of quantum Fisher information (QFI), an essential quantity for
characterizing quantum sensors. Here we investigate multi-mode quadratic
bosonic systems, which exhibit higher-order EP dynamics, but possess
Hermitian Hamiltonians without Langevin noise, thus can be utilized for
quantum sensing. We derive an exact analytic formula for the QFI, from which
we establish a scaling relation between the QFI and the order of the EP. We
apply the formula to study a three-mode EP sensor and a multi-mode bosonic
Kitaev chain and show that the EP physics can significantly enhance the
sensing sensitivity. Our work establishes the connection between two
important fields: non-Hermitian EP dynamics and quantum sensing, and may
find important applications in quantum information and quantum non-Hermitian
physics.
\end{abstract}

\maketitle

\paragraph{Introduction.\thinspace ---}

Exceptional points (EPs), the degenerate points of non-Hermitian systems
where two or more eigenstates coalesce \cite%
{Bender1998real,AshidaGongUeda,BBK2021rmp,Ozdemire2019}, have recently
emerged as a novel platform for achieving high precision sensing of physical
parameters \cite{Wiersig2014enchanced, feng2014single,
hodaei2017enhanced,lau2018fundamental,luo2022quantum,wang2019non-hermitian,mcdonald2020exponentially,chen2017exceptional,Budich2020sensor,PhysRevA.98.023805,Zhang2019quantum,chen2019sensitivity, Liu2016metrology,Lin2016enhanced,kononchuk2022exceptional,hodaei2014parity,PTenhance1}%
. In a classical system, the high sensitivity stems from the scaling of the
eigenspectrum $\sim \omega ^{1/2}$ of a typical second-order (\textit{i.e.},
two-fold degeneracy) EP, leading to a divergent spectrum response rate $%
d\left( \Delta \omega \right) /d\epsilon \propto \epsilon ^{-1/2}$ under a
perturbation $\epsilon $ of a physical parameter deviated from the EP \cite%
{Wiersig2014enchanced,chen2017exceptional}. For a higher ($M\geq 3$) order
EP with $\Delta \omega \sim \epsilon ^{d_{\omega }}$ ($d_{\omega }\geq 1/M$%
), the divergence can be more substantial $\sim \epsilon ^{1/M-1}$ to
achieve higher sensitivity \cite%
{hodaei2017enhanced,Lin2016enhanced,PTenhance1}.

While the EP-based sensing is well studied in classical open systems like
gain/loss nanophotonics, its generalization to a quantum system poses a
fundamental challenge \cite%
{PhysRevA.98.023805,Zhang2019quantum,lau2018fundamental,
Choi2017,Chu2020,Naghiloo2019,Wu2019,Yu2020,Scheel2018}. In an open quantum
system, the intrinsic Langevin noises may break the underlying symmetry
(e.g., parity-time) that protects the EP, rendering the conceptual
difficulty for even defining EP-based quantum sensor \cite{Scheel2018}.
Recently, it was shown in a two-mode bosonic parametric amplification
process, an EP could emerge from the dynamical evolution matrix of the
Hermitian quadratic bosonic Hamiltonian, thus avoids Langevin noise \cite%
{wang2019non-hermitian,luo2022quantum}. Around the EP, the quantum dynamics
are very sensitive to the small perturbation of the parameter, therefore can
be utilized as a quantum EP sensor \cite{luo2022quantum}.

The emergence of the quantum EP in the two-mode bosonic Hamiltonian and its
application in quantum sensing naturally raise questions about the general
EP physics and the characterization of EP-based quantum sensors in
multi-mode quadratic bosonic Hamiltonians, which have been experimentally
engineered in nonlinear optical media \cite{Wang2017,Zhang2020},
multi-frequency superconducting parametric cavities \cite%
{Busnaina2023,Gaikwad2023}, and optomechanical systems \cite%
{Wanjura2023,Pino2022,Slim2023} in recent years. In quantum sensing, quantum
Fisher information (QFI) is one main characteristic quantity that provides a
low bound for sensing precision through quantum Cram\'{e}r-Rao bound \cite%
{Rmp1,Rmp2}. QFI is very different from the divergence of the spectrum
response rate that characterizes classical EP-based sensors. It is unclear
how the scaling of the quantum EP spectrum is connected with the behavior of
the QFI and whether there exists certain universal scaling of the QFI
at/around the EP.

To address these questions, we study the QFI in a generic multi-mode bosonic
quadratic Hamiltonian, which is Hermitian but its Heisenberg equation of
motion (HEOM) is governed by a non-Hermitian dynamical matrix \cite%
{Lieu2018topological,McDonald2018phase,Squeezingpg}, yielding the EP
physics.\ Our main results are:

\textit{i}) We derive an analytic formula of the QFI that is generally hard to
calculate even numerically for multi-mode quadratic Hamiltonians due to the exponentially large Hilbert space of particle numbers. 
Current methods for the estimate of QFI relies on approximate the input-out theory in quantum optics \cite{mcdonald2020exponentially,RMP3},
instead of the direct evaluation over the quantum wavefunction.      
We further explore of the 
applications of the analytic formula in a three-mode quantum sensor as well as the multi-mode
bosonic Kitaev chain.

\textit{ii}) Utilizing the analytic formula, we find a universal scaling of
the QFI $F(t)\sim t^{d_{F}}$ with $d_{F}\leq 4M-2$ at an $M$-th order EP for
a large time $t$, which establishes the connection between the QFI and the
order of the EP. From the analytic formula and scaling relation, we show: 1)
EP can significantly enhance the sensitivity: the achievable $t^{4M-2}$
scaling showcases the QFI can grow much faster over time than non-EP sensors
with typically $\sim t^{2}$ scaling; 2) Different from the divergence of the
spectrum response rate, there is no divergence of the QFI at EP due to the
continuity of the QFI formula, therefore the large EP sensitivity
enhancement can be achieved for parameters close to the EP.

\textit{iii}) In an $N$-mode bosonic Kitaev chain, near the EP the QFI per
particle at a fixed time has an exponential scaling $F(t)\sim \beta ^{2N-2}$
($\beta >1$ depends on parameters of the Kitaev chain), which indicates the
sensitivity of quantum EP sensor can increase exponentially with the system
size, paving the way for designing novel quantum EP sensors through
size/mode engineering.


\paragraph{Analytic formula for QFI:}

\begin{figure}[t]
\centerline{\includegraphics[width=3.2in]{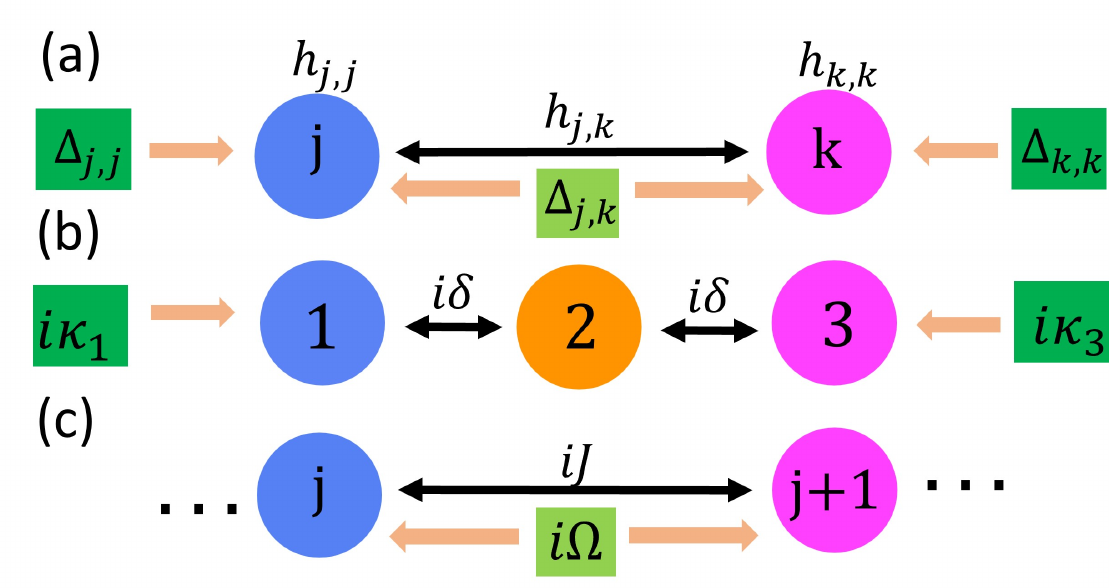}}
\caption{(a) Schematic diagram of different terms in the generic Hamiltonian
(\protect\ref{Ha}). (b) and (c) illustrate the three mode model and bosonic
Kitaev chain, respectively.}
\label{F1}
\end{figure}
Consider a generic $N$-mode Hermitian bosonic quadratic Hamiltonian (\textit{%
i.e.}, Bogoliubov Hamiltonian), 
\begin{equation}
\hat{H}=\sum_{j,k=1}^{N}(h_{j,k}\hat{a}_{j}^{\dagger }\hat{a}_{k}+\frac{%
\Delta _{j,k}}{2}\hat{a}_{j}^{\dagger }\hat{a}_{k}^{\dagger }+\frac{\Delta
_{j,k}^{\ast }}{2}\hat{a}_{j}\hat{a}_{k}),  \label{Ha}
\end{equation}%
where $\hat{a}_{k}$ is the bosonic annihilation operator, $%
h_{j,k}=h_{k,j}^{\ast }$, $\Delta _{j,k}=\Delta _{k,j}$ due to the
Hermiticity of $\hat{H}$ and bosonic commutation relation. A schematic
diagram of different terms in the Hamiltonian is shown in Fig. \ref{F1}(a).
The Heisenberg equation of motion (HEOM) is $i\frac{d}{dt}\hat{V}(t)=H_{D}%
\hat{V}(t)$, where $\hat{V}(t)=\left[ 
\begin{array}{cc}
A, & A^{\dagger }%
\end{array}%
\right] ^{T}$, $A=\left( \hat{a}_{1},...,\hat{a}_{N}\right) $, $T$ is the
transpose operator, 
\begin{equation}
H_{D}=\left( 
\begin{array}{cc}
h & \Delta \\ 
-\Delta ^{\ast } & -h^{\ast }%
\end{array}%
\right)  \label{DM1}
\end{equation}%
is the dynamical matrix, $h$ and $\Delta $ are $N\times N$ matrices. $H_{D}$
satisfies the symmetries $\tau _{x}H_{D}\tau _{x}=-H_{D}^{\ast }$, $\tau
_{z}H_{D}\tau _{z}=H_{D}^{\dagger }$ and $\tau _{y}H_{D}\tau _{y}=-H_{D}^{%
\mathbf{T}}$ \cite%
{bernard2002classification,KawabataShiozakiUedaSato,ZhouLee,DefectsLiuChen,Liu2022symmetry}%
, where $\tau _{x},\tau _{y}$, and $\tau _{z}$ are Pauli matrices. According
to HEOM, we have $\hat{V}(t)=e^{-iH_{D}t}\hat{V}(0)=S(t)\hat{V}(0)$.

In terms of the real and imaginary parts of $h$ and $\Delta $, the dynamical
matrix can be written as 
\begin{equation}
H_{D}=\tau _{0}\otimes ih_{I}+\tau _{z}\otimes h_{R}+\tau _{x}\otimes
i\Delta _{I}+i\tau _{y}\otimes \Delta _{R},  \label{DM2}
\end{equation}%
where $h_{I}^{T}=-h_{I}$, $h_{R}^{T}=h_{R}$, $\Delta _{I}^{T}=\Delta _{I}$, $%
\Delta _{R}^{T}=\Delta _{R}$ are all real matrices. $H_{D}$ generally does
not possess EPs (e.g., $\Delta =0$ and $H_{D}$ is Hermitian), and the EPs
may exist when both $h$ and $\Delta ~$are nonzero. In the presence of
certain symmetry of $H_{D}$, the $2N\times 2N$ dynamical matrix can be block
diagonalized into two $N\times N$ matrices. For instance, when $h_{R}=\Delta
_{R}=0$ (e.g., the three-mode and Kitaev chain shown below), $H_{D}$
satisfies a symmetry $\left[ H_{D},\tau _{x}\right] =0$, allowing the block
diagonalization of $H_{D}$. In each block, the EP is determined by the
corresponding block matrix, but the maximum order of EPs $M\leq N$.

We consider a $N$-mode coherent initial state given by $|\psi _{0}\rangle
=|\alpha _{1},...,\alpha _{N}\rangle =\hat{D}_{1}(\alpha _{1})...\hat{D}%
_{N}(\alpha _{N})|0\rangle $ with $\hat{D}_{j}(\alpha _{j})=e^{\alpha _{j}%
\hat{a}_{j}^{\dagger }-\alpha _{j}^{\ast }\hat{a}_{j}}$. The quantum state
at time $t$ becomes $|\psi _{t}\rangle =e^{-i\hat{H}t}|\psi _{0}\rangle $.
With a tedious calculation, we find the QFI $F_{\eta }=4[\langle \partial
_{\eta }\psi _{t}|\partial _{\eta }\psi _{t}\rangle -|\langle \psi
_{t}|\partial _{\eta }\psi _{t}\rangle |^{2}]$ for a sensing parameter $\eta 
$ is \cite{SM} 
\begin{equation}
F_{\eta }=4\bm{B}^{\dagger }\bm{B}+2\mathrm{Tr}[\bm{C}_{2}^{\dagger }\bm{C}%
_{2}],  \label{QFIcm}
\end{equation}%
where $\bm{B}=\bm{C}_{1}\bm{\alpha}+\bm{C}_{2}\bm{\alpha}^{\ast }$, $%
\bm{\alpha}=[\alpha _{1},...,\alpha _{N}]^{\mathrm{T}}$ and $\bm{C}_{1}$ and 
$\bm{C}_{2}$ are $N\times N$ matrices given by 
\begin{equation}
\left( 
\begin{array}{cc}
\bm{C}_{1} & \bm{C}_{2} \\ 
\bm{C}_{2}^{\ast } & \bm{C}_{1}^{\ast }%
\end{array}%
\right) =\int_{y=0}^{t}dy[S(y)]^{\dagger }\Sigma _{z}\partial _{\eta
}H_{D}S(y),  \label{c1c2}
\end{equation}%
where $\Sigma _{z}=\tau _{z}\otimes \mathbb{I}$, $S(y)=e^{-iyH_{D}}$, $%
\mathbb{I}$ is the $N\times N$ identity matrix.

Eqs. (\ref{QFIcm}) and (\ref{c1c2}) also apply to time-dependent systems
with $S(y)=\mathcal{T}[exp({-i\int_{x=0}^{y}dxH_{D}(x)})]$ for
time-dependent dynamical matrix $H_{D}(t)$, where $\mathcal{T}$ is the time
ordering operator. The QFI formula for other initial states beyond coherent
states is given in the supplementary materials \cite{SM}.

The scaling of the QFI with time can be derived from Eq. (\ref{QFIcm}). At
an $M$-th order EP, when the imaginary parts of the eigenvalues of $H_{D}$
are equal to zero (i.e., stable region), there always exists at least one
matrix element in $S(y)=e^{-iyH_{D}}$, whose leading order is proportional
to $y^{M-1}$ \cite{AshidaGongUeda}. According to Eq. (\ref{c1c2}), the
leading matrix elements of $\bm{C}_{1}$ and $\bm{C}_{2}$ are proportional to 
$t^{2M-1}$ except for accidental parameter regions where the coefficient
cancellation leads to lower order of \thinspace $t$ dependence. Hence $%
F_{\eta }(t)\sim t^{d_{F}}$ with $d_{F}\leq 4M-2$, and the maximum $%
d_{F}=4M-2$ can be achieved for general parameter regions. In contrast, for
a Hermitian Hamiltonian, $S_{y}$ is purely a phase and $C_{j}\sim t$, thus
the QFI is $\sim t^{2}$.

The above results can be better illustrated with a simple single mode ($N=1$%
) Hamiltonian 
\begin{equation}
\hat{H}_{1}=\delta \hat{a}^{\dagger }\hat{a}+i\frac{\kappa }{2}(\hat{a}%
^{\dagger 2}-\hat{a}^{2}),  \label{Hsm}
\end{equation}%
where the dynamical matrix $H_{D1}=\delta \tau _{z}+i\kappa \tau _{x}$. $%
\kappa $ is the single mode squeezing parameter, and $\delta $ is the phase
mismatch. $H_{D1}$ obeys the anti-$\mathcal{PT}$ symmetry $\left\{ H_{D1},%
\mathcal{PT}\right\} =0$ with $\mathcal{P}=\tau _{x}$ and $\mathcal{T=K}$,
and possesses a second order ($M=2$) EP at $\delta =\kappa $. The energy
spectrum response $\Delta \omega _{\pm }=\pm \left( \delta \epsilon \right)
^{1/2}$ for a perturbation $\epsilon =\delta -\kappa $ from the EP for the
parameter $\kappa $. At the EP, the time evolution matrix $S(t)=\left( 
\begin{array}{cc}
1-it\delta & t\delta \\ 
t\delta & 1+it\delta%
\end{array}%
\right) $ with the leading order $\sim t^{M-1}=t$. For the sensing parameter 
$\kappa $, $\partial _{\kappa }H_{D1}=i\tau _{x}$, $\Sigma _{z}=\tau _{z}$,
and the resulting $C_{1}=-4\delta ^{2}t^{3}/3$ and $C_{2}=-2i\delta
^{2}t^{3}/3+\delta ^{2}t^{2}+it$, with the leading order $\sim
t^{2M-1}=t^{3} $. Therefore the QFI scales as $F_{\kappa }=4|C_{1}\alpha
+C_{2}\alpha ^{\ast }|^{2}+2|C_{2}|^{2}\sim t^{6}$ with $d_{F}=$ $4M-2=6$ as
predicted by the scaling law. Hereafter, we apply the analytic formula and
the scaling law to two complex multi-mode examples: a three mode sensor and
a multi-mode Kitaev chain.

\paragraph{Three mode quantum EP sensor.\thinspace ---}

Three coupled optical cavities with gain/neutral/loss pattern exhibit a
third order EP with enhanced sensitivity, as demonstrated in recent
experiments \cite{hodaei2017enhanced}. Here we consider a similar three mode
Hamiltonian%
\begin{equation}
\hat{H}_{3}=i\delta \hat{a}_{1}^{\dagger }\hat{a}_{2}+i\delta \hat{a}%
_{2}^{\dagger }\hat{a}_{3}+\frac{i\kappa _{1}}{2}\hat{a}_{1}^{2}+\frac{%
i\kappa _{3}}{2}\hat{a}_{3}^{\dagger 2}+\mathrm{h.c.,}  \label{Htsm}
\end{equation}%
where the gain/loss pattern is replaced by the single mode squeezing in
modes 1 and 3, as illustrated in Fig. \ref{F1}(b). The dynamical matrix can
be written as $H_{D3}=\tau _{0}\otimes \mathbf{K}_{1}+\tau _{x}\otimes 
\mathbf{K}_{2}$ with $\mathbf{K}_{1}=-\delta (\lambda _{2}+\lambda _{7})$, $%
\mathbf{K}_{2}=\mathrm{diag}[-i\kappa _{1},0,i\kappa _{3}]$, and $\lambda
_{2}$ and $\lambda _{7}$ being Gell-Mann matrices \cite%
{Gell-Mann1962symmetries}. As discussed previously, the $\tau_{x}$ symmetry
allows the diagonalization of $H_{D3}$ into two irreducible blocks $H_{D3\pm
}=\mathbf{K}_{1}\pm \mathbf{K}_{2}$. $H_{D3\pm }$ can be regarded as the
quantum generalization of the gain/neutral/loss cavity model \cite%
{hodaei2017enhanced}.

\begin{figure}[t]
\centerline{\includegraphics[width=3.5in]{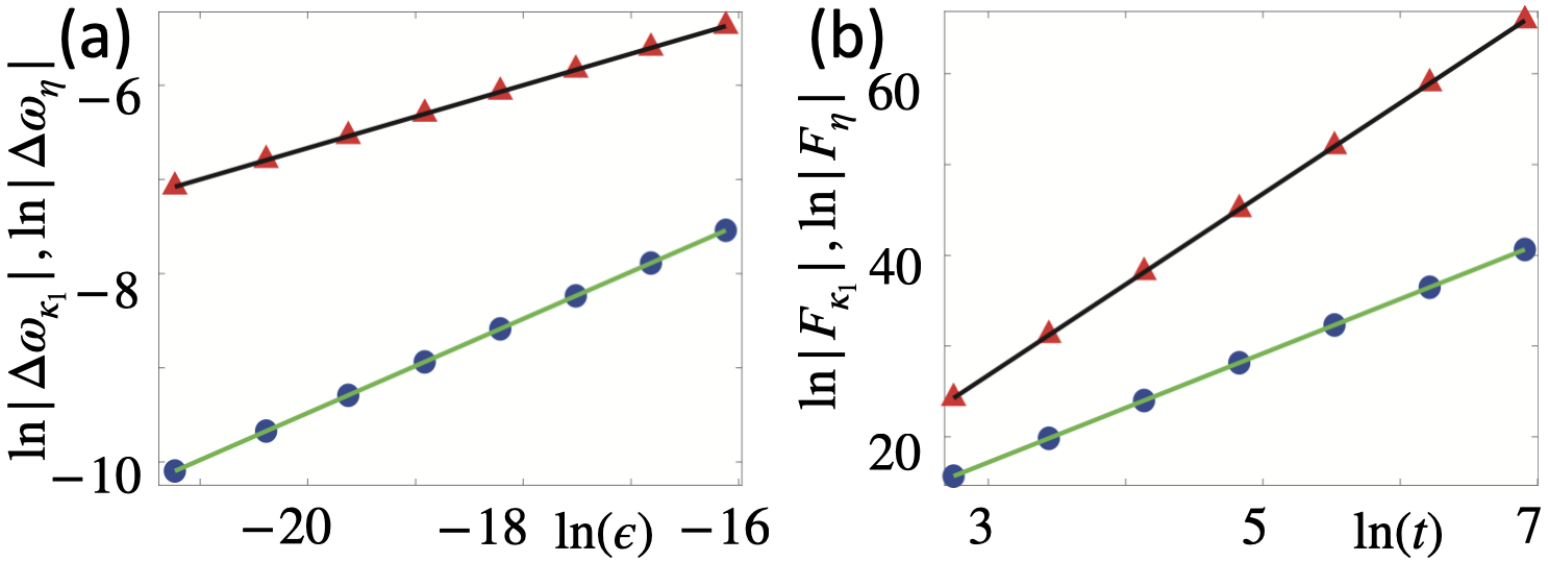}}
\caption{Maximum spectrum response and QFI for the three mode model.
Parameters $\protect\delta =1$ and $\protect\kappa _{1}=\protect\kappa _{3}=%
\protect\sqrt{2}$ for the EP. (a) Plot of maximum spectrum responses versus
the perturbation $\protect\epsilon $. (b) Plot of QFI versus time $t$. In
(a) and (b), the red triangles and blue dots are numerical results without
and with constraint $\protect\kappa _{1}=\protect\kappa _{3}=\protect\eta $,
respectively. The solid lines are fitting functions.}
\label{F2}
\end{figure}

Each block exhibits a third-order ($M=3$) EP at $\sqrt{2}\delta =\kappa
_{1}=\kappa _{3}$ with three degenerate eigenvalues $\omega _{k}^{\pm }=0$
for $k=1,2,3$. Consider a perturbation of $\kappa _{1}=\sqrt{2}\delta
+\epsilon $ ($\epsilon <<\delta $) from the EP for mode 1, the spectrum response
 $\Delta \omega _{k}^{\pm }=\omega _{k}^{\pm }\left( \epsilon \right)
-\omega _{k}^{\pm }\left( 0\right) =ie^{\pm i\pi k/3}\delta ^{2/3}\epsilon
^{1/3}$ at the leading order, showing a $\epsilon ^{1/3}$ scaling behavior
(i.e., the exponent $d_{\omega }=1/3$) \cite{SM}, similar as that for
classical coupled cavities \cite{hodaei2017enhanced}. In Fig. \ref{F2}(a),
we plot the numerical results for the maximum spectrum response $\Delta
\omega _{\kappa _{1}}=\mathrm{max}_{\pm,k}|\Delta \omega _{k}^{\pm}|$ with
respect to $\epsilon $ (red triangles) and the linear fitting $\ln (\Delta
\omega _{\kappa _{1}})=0.33\ln (\epsilon )+0.01$ (black solid line), showing
excellent agreement with the analytic result.

For quantum sensing of the parameter $\kappa _{1}$, we find $F_{\kappa
_{1}}\propto t^{10}$ when $t\rightarrow +\infty $ from Eq. (\ref{QFIcm}) 
\cite{SM} for a coherent initial state. In Fig. \ref{F2}(b), we show the
numerical result of $\ln (F_{\kappa _{1}})$ as a function of $\ln (t)$, and
its linear fitting $\ln (F_{\kappa _{1}})=10\ln (t)-3.22$, showing the
numerical result agrees well with our analytical prediction \cite{SM}. $%
d_{F}=4M-2=10$ reaches the maximum exponent of the scaling. Far from the EP
with $\kappa _{1}=\kappa _{3}=0$, $F_{\kappa _{1}}\propto t^{2}$ when $%
t\rightarrow +\infty $ \cite{SM}, similar as that for a Hermitian sensor.
The $t^{2}$ to $t^{10}$ scaling change of the QFI demonstrates the
EP-enhanced sensitivity.

Different perturbations for the same EP may induce different spectrum
responses, leading to different scalings of the QFI. In the three mode
Hamiltonian (\ref{Htsm}), a constraint $\kappa _{1}=\kappa _{3}=\eta $ with
a perturbation $\eta =\sqrt{2}\delta +\epsilon $ for both modes 1 and 3
around the third order EP $\eta =\sqrt{2}\delta $ leads to the double
degenerate spectrum response $\Delta \omega _{1}^{\pm}=0$, $\Delta \omega _{
2}^{\pm}= i\sqrt{2\sqrt{2}\delta \epsilon }$, and $\Delta \omega _{
3}^{\pm}=- i\sqrt{2\sqrt{2}\delta \epsilon }$. They have the $\epsilon ^{1/2}$
scaling, agreeing with the numerical result and its fitting line $\ln
(\Delta \omega _{\eta})=0.5\ln (\epsilon )+0.51$ in Fig. \ref{F2}(a), where $%
\Delta \omega _{\eta}=\mathrm{max}_{\pm,k}|\Delta \omega _{k}^{\pm}|$ for
the perturbation in $\eta$. The QFI $F_{\eta }\propto t^{6}$ when $%
t\rightarrow +\infty $ \cite{SM}, which also shows excellent agreement with
the numerical result and its linear fitting $\ln (F_{\eta })=6\ln (t)-0.82$
in Fig. \ref{F2}(b). Here $d_{F}=6<4M-2$ and $d_{\omega }=1/2>1/M$.


\paragraph{Bosonic Kitaev chain\,---}

We extend the analysis from three modes to $N$ modes and consider the
bosonic Kitaev chain with the Hamiltonian \cite%
{kitaev2001unpaired,McDonald2018phase,mcdonald2020exponentially} 
\begin{equation}
\hat{H}_{BKC}=\sum_{j=1}^{N-1}\left( iJ\hat{a}_{j}^{\dagger }\hat{a}%
_{j+1}+i\Omega \hat{a}_{j}^{\dagger }\hat{a}_{j+1}^{\dagger }+\mathrm{h.c.}%
\right) .
\end{equation}%
The dynamical matrix for the HEOM $H_{DK}=\tau _{0}\otimes \mathbf{L}%
_{1}+\tau _{x}\otimes \mathbf{L}_{2}$, where $\mathbf{L}_{1}$ and $\mathbf{L}%
_{2}$ are $N\times N$ matrices with $(\mathbf{L}_{1})_{j,j+1}=-(\mathbf{L}%
_{1})_{j+1,j}=iJ$ and $(\mathbf{L}_{2})_{j,j+1}=(\mathbf{L}%
_{2})_{j+1,j}=i\Omega $ (all other elements are $0$). The model is
schematically illustrated in Fig. \ref{F1}(c). Under unitary transformation $%
(\tau _{x},\tau _{y},\tau _{z})\rightarrow (\tau _{z},\tau _{y},-\tau _{x})$%
, $H_{DK}$ can be block diagonalized as $\tau _{0}\otimes \mathbf{L}%
_{1}+\tau _{z}\otimes \mathbf{L}_{2}$ with $H_{DK+}=\mathbf{L}_{1}+\mathbf{L}%
_{2}$ and $H_{DK-}=\mathbf{L}_{1}-\mathbf{L}_{2}$ being two irreducible
blocks that represent Hatano-Nelson model \cite{Hatano1996localization}. At $%
J=\Omega $, both blocks $H_{DK\pm }$ reduce to the $N$-dimensional Jordan
normal forms with zero eigenvalues in the diagonal, yielding two degenerate $%
N$-th order (\textit{i.e.}, $M=N$) EPs.

A perturbation from the EP $\epsilon =J-$ $\Omega $ leads to the change of
the $2N$ eigenvalues 
\begin{equation}
\Delta \omega _{k}=2\sqrt{(2J-\epsilon )\epsilon }\cos (\frac{k\pi }{N+1}%
)\quad
\end{equation}%
for $k=1,...,N$, which are double degenerate for two blocks. Clearly, $%
\Delta \omega _{k}=0$ for $k=(N+1)/2$ with an odd $N$, while all other
spectrum responses $\Delta \omega _{k}\propto \epsilon ^{1/2}$ do not reach
the maximum scaling $\epsilon ^{1/N}$. The QFI $F_{\Omega }\propto t^{4j+2}$
for both $N=2j$ and $2j+1$, indicating each non-zero $\sqrt{\epsilon }$
spectrum may contribute $t^{2}$ scaling of QFI. The scaling factor $%
d_{F}=4j+2<d_{F}^{\max }=4M-2$.

\begin{figure}[t]
\centerline{\includegraphics[width=3.5in]{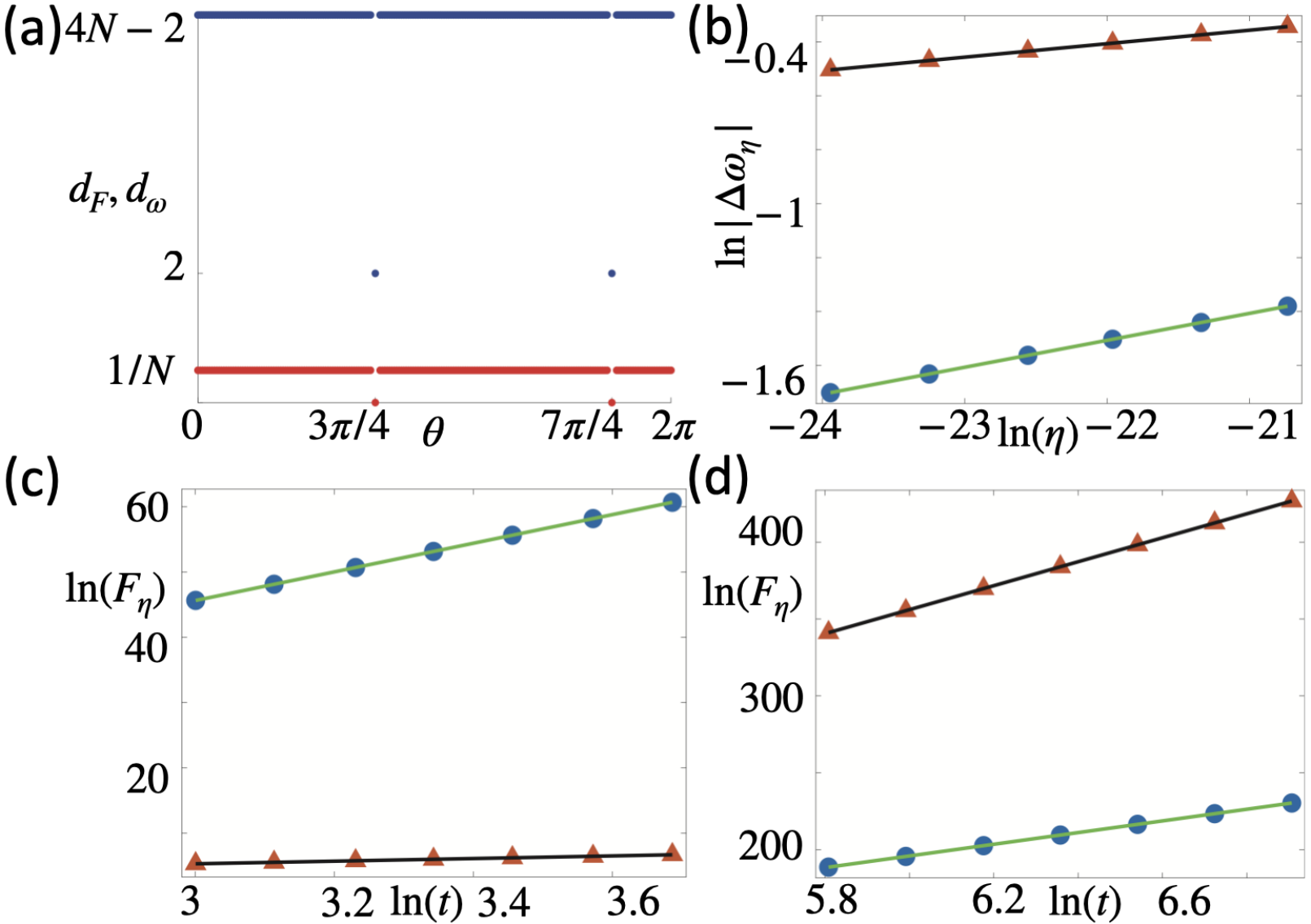}}
\caption{Maximum spectrumn response and QFI of bosonic Kitaev chain at $%
J=\Omega =1$. (a) Schematic diagram of the exponents $d_{F}$ (blue) and $d_{%
\protect\omega }$ (red) versus $\protect\theta $. (b) Maximum spectrum
response versus $\ln (\protect\eta )$. The red triangles (blue dots) are
numerical results of $\ln (\Delta \protect\omega _{\protect\eta })$ for $%
N=20 $ $(10)$. (c) The red triangles (blue dots) are numerical results of
QFI versus $\ln (t)$ for $\protect\theta =3\protect\pi /4$ ($\protect\theta %
=2.99\protect\pi /4$). $\protect\eta =0$ and $N=6$. (d) QFI versus $\ln (t)$%
. The red triangles (blue dots) are numerical results of $\ln (F_{\protect%
\eta })$ for $N=20$ $(10)$. The solid lines in (b-d) are fitting functions.}
\label{F3}
\end{figure}

In order to reach the maximum\ spectrum response exponent $d_{\omega }=1/N$,
we consider a local perturbation 
\begin{equation}
\hat{H}_{\eta }=i\eta \left[ \sin (\theta )\hat{a}_{N}^{\dagger }\hat{a}%
_{1}+\cos (\theta )\hat{a}_{N}^{\dagger }\hat{a}_{1}^{\dagger }\right] +%
\mathrm{h.c.}
\end{equation}%
that couples modes 1 and $N$. $\hat{H}_{\eta }$ adds $(\mathbf{L}%
_{1})_{N,1}=-(\mathbf{L}_{1})_{1,N}=i\eta \sin (\theta )$ and $(\mathbf{L}%
_{2})_{N,1}=(\mathbf{L}_{2})_{1,N}=i\eta \cos (\theta )$ in the dynamical
matrix. We find the maximum spectrum response $\Delta \omega _{\eta }=%
\mathrm{max}_{k}(|\Delta \omega _{k}|)\propto \eta ^{1/N}$ (i.e., $d_{\omega
}=1/N$) for $\theta \neq \frac{3\pi }{4}$ and $\frac{7\pi }{4}$, while $%
\Delta \omega _{\eta }=0$ (i.e., $d_{\omega }=0$) for $\theta =\frac{3\pi }{4%
}$ or $\frac{7\pi }{4}$ \cite{SM}. From Eq. (\ref{QFIcm}), we find $F_{\eta
}\propto t^{4N-2}$ (i.e., $d_{F}=4N-2$) for $\theta \neq \frac{3\pi }{4}$
and $\frac{7\pi }{4}$, and $F_{\eta }\propto t^{2}$ (i.e., $d_{F}=2$) for $%
\theta =\frac{3\pi }{4}$ or $\frac{7\pi }{4}$ \cite{SM} when $t\rightarrow
\infty $. In Fig. \ref{F3}(a), we illustrate $d_{F}$ and $d_{\omega }$ as a
function of $\theta $ at the EP $J=\Omega =1$ and $\eta =0$.

The analytical results are confirmed by the numerical calculations. In Fig. %
\ref{F3}(b), we plot the numerical maximum spectrum response with respect to
the perturbation $\eta $ around the EP $J=\Omega =1$ for the system size $%
N=20$ and $10$), together with their fittings $\ln (\Delta \omega _{\eta
})=0.05\ln (\eta )+0.69$ and $\ln (\Delta \omega _{\eta })=0.1\ln (\eta
)+0.69$. They agree well with the analytical expressions for the response
exponent $1/N$ shown in Fig. \ref{F3}(a). In Fig. \ref{F3}(c), we plot $%
\ln (F_{\eta })$ versus $\ln (t)$ for $N=6$ at the EP with $\eta =0$ for
slightly different $\theta =3\pi /4$ and $\theta =2.99\pi /4$. The fitting
functions are found to be $\ln (\Delta \omega _{\eta })=2.02\ln (t)-0.74$
and $\ln (\Delta \omega _{\eta })=21.92\ln (t)-20.14$, respectively. The
result confirms the $t^{2}$ scaling of the QFI at the accidental degenerate
point $\theta =3\pi /4 $. In Fig. \ref{F3}(d), we plot $\ln (F_{\eta })$
versus $\ln (t)$ at the EP for two different system sizes $N=20$ and $10$,
with their linear fitting functions $\ln (F_{\eta })=77.93\ln (t)-111.49$
and $\ln (F_{\eta })=37.97\ln (t)-31.89$. We see they satisfy $d_{F}=4N-2$,
reaching the maximum scaling exponent for the QFI.

Far from the EP with $J\neq 0$ and $\eta =\Omega =0$, we find $F_{\eta
}\propto t^{2}$ when $t\rightarrow +\infty $ from Eq. (\ref{QFIcm}).
Therefore our analytical and numerical results demonstrate the enhanced
quantum sensitivity at the EP (from $t^{2}$ to $t^{4N-2}$). The Kitaev chain
model also indicates that the maximum scaling $t^{4M-2}$ can be reached for
an $M$-th order EP when the dynamical matrix can transform to the Jordan
normal form.

\begin{figure}[t]
\centerline{\includegraphics[width=3.5in]{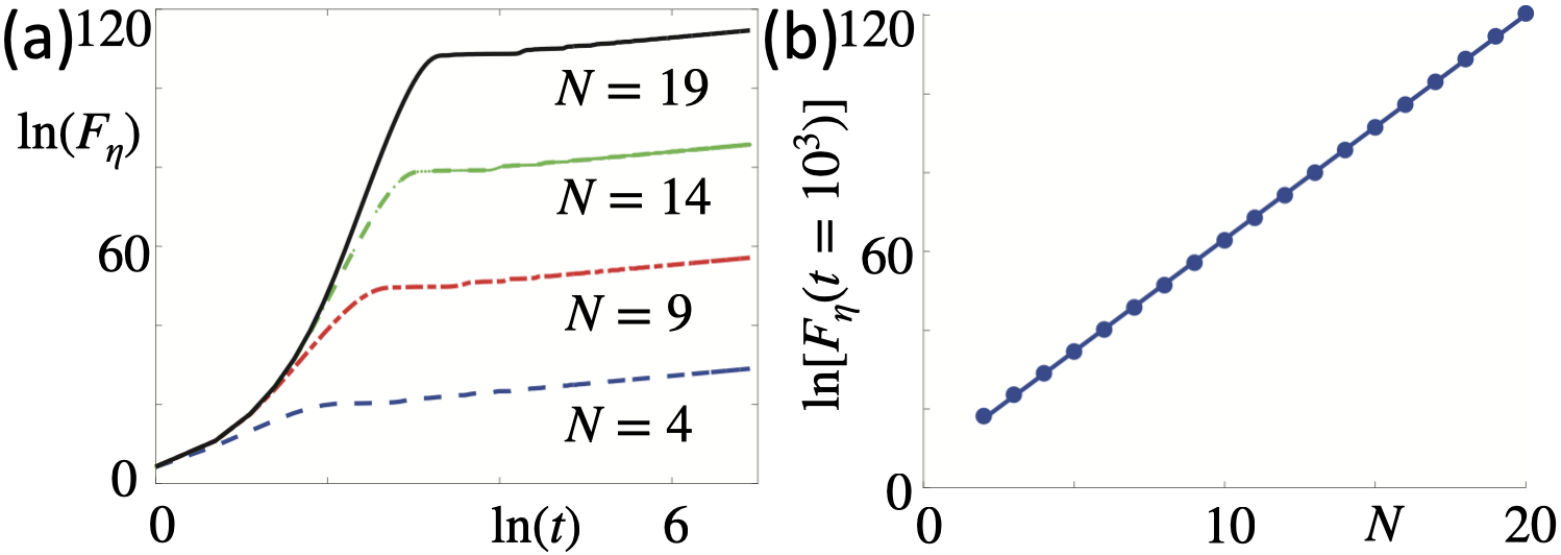}}
\caption{The parameters are $J=1$, $\Omega =0.9$, and $\protect\eta =0$. (a)
QFI $\ln (F_{\protect\eta })$ versus $\ln (t)$ for $N=4,9,14,19$.
(b) Fixed time logarithm QFI $\ln [F_{\protect\eta }(t_{0}=1000)]$ versus size $%
N $. The blue dots are numerical results and the solid line is the
fitting function.}
\label{F4}
\end{figure}

At an $N$-th order EP and with a fixed large time $t=t_{0}$, $F_{\eta }$ is
proportional to $t_{0}^{4N-2}$ at the leading order, indicating the
sensitivity may be exponentially enhanced by the system size. However, such $%
t_{0}$ may approach infinity exactly at the EP. Since the QFI is a
continuous function of time and other parameters, the enhanced sensitivity
at the EP may pass to the nearby interval around the EP, where $t_{0}$ is
finite. We consider the parameters near the EP with $\eta =0$, $\beta =\sqrt{%
\frac{J+\Omega }{J-\Omega }}\gg 1$ and a fixed but finite large time $t_{0}$%
. We can diagonalize the dynamical matrix $H_{DK}$ with a unitary matrix and
find that $F_{\eta }(t=t_{0})$ is proportional to $\beta ^{4N-4}$ at the
leading order, \textit{i.e.}, $\ln (F_{\eta }(t=t_{0}))$ is approximately
proportional to $4N\ln (\beta )$.

In Fig. \ref{F4}(a), we plot the numerical calculation of $\ln (F_{\eta })$
as a function of $\ln (t)$ for $N=4,9,14,19$, with parameters $J=1$, $\Omega
=0.9$, $\theta =\pi /4$, and $\eta =0$. We see the growth of QFI with time
becomes polynomial after certain time, indicating the saturation of the
initial exponential growth due to the finite size of the system. In fact, in
a periodic Kitaev chain, the parameter $J=1$ and $\Omega =0.9$ yields
imaginary eigenspectrum, leading to the exponential growth of the QFI with
time. However, the finite size spectrum is still real, leading to the
saturation of QFI after the boundary effect kicks in. In Fig. \ref{F4}(b),
we show the numerical calculation of $\ln (F_{\eta })$ with respect to $N$
at a fixed time $t_{0}=1000$, with the fitting equation $\ln [F_{\eta
}(t=1000)]=5.69N+6.17$. The numerical coefficient $5.69$ is consistent with
theoretical coefficient  $4\ln (\beta )\approx 5.89$.

Finally, the leading term of total particle number $Q=\langle \psi
_{t}|\sum_{j=1}^{N}\hat{a}_{j}^{\dagger }\hat{a}_{j}|\psi _{t}\rangle $
scales as $Q\sim \beta ^{2N-2}$, therefore the QFI at the fixed time reaches
the Heisenberg limit with $F_{\eta }\sim Q^{2}$, indicating the QFI per
particle is exponentially enhanced. Such exponentially enhanced sensitivity
may originate from the non-Hermitian skin effect (NHSE). In the bosonic
Kitaev chain, there is an NHSE for $H_{DK}$ without the perturbation $\eta $%
, leading to the accumulation of the wavefunction at the boundary. However,
the boundary coupling from the perturbation $\eta $ leads to the
disappearance of the NHSE, which causes a dramatic change of the
wavefunction and energy spectrum, yielding the exponential enhancement of
the QFI per particle \cite{Budich2020sensor}. This reveals that previous
conclusion that NHSE cannot be used to enhance per-particle sensitivity \cite%
{mcdonald2020exponentially} based on the input-out theory may be model dependent.


\paragraph{Conclusion and discussion.\thinspace ---}

In this Letter, we derive the exact QFI formula for a generic quadratic
Bosonic Hamiltonian. Utilizing the exact formula, we establish the
connection between the QFI scaling exponent and the order of EP of the
dynamical matrix. Our analytic methods can also be applied to other
important models, e.g., quadratic bosonic Hamiltonian with topological bands 
\cite{Budich2020sensor,Squeezingpg}, for studying quantum squeezing,
entanglement and sensing of topological multi-mode chains. While our work on the
QFI establishes the lower bound for the ultimate precision for quantum EP
sensors, the optimal strategies for achieving such bound are unknown and
demand developing well-designed measurement schemes \cite{Zhang2021}. Our
result bridges the fields of quantum sensing and non-Hermitian EP physics
and may be useful for the design of new EP-based quantum sensors.

\begin{acknowledgments}
\emph{Acknowledgement:} We thank the helpful discussions with Q. Zhuang and
X. Zhao. C.L. and C.Z. are supported by AFOSR (FA9550-20-1-0220) and NSF
(PHY-2409943, OMR-2228725, ECC-2411394). J.W. and D.S. are supported by DOE
(DE-SC0022069). J.W. is also support by NSF (OMR-2329027). L.Y. acknowledges support from 
the McKelvey School of Engineering at Washington University in St. Louis.
\end{acknowledgments}


\widetext

\renewcommand{\theequation}{S\arabic{equation}} \renewcommand{\thefigure}{S%
\arabic{figure}} \renewcommand{\thetable}{S\arabic{table}} 
\setcounter{equation}{0} \setcounter{figure}{0} \setcounter{table}{0} 


\section{SI.~Derivation of the analytic formula for quantum Fisher
information}

\subsection{A. QFI for coherent initial state and time-independent
Hamiltonian}

In this section, we derive the quantum Fisher information (QFI) formula for
a sensing parameter $\eta $ in a generic multimode bosonic quadratic
Hamiltonian 
\begin{equation}
\hat{H}=\sum_{j,k=1}^{N}(h_{j,k}\hat{a}_{j}^{\dagger }\hat{a}_{k}+\frac{%
\Delta _{j,k}}{2}\hat{a}_{j}^{\dagger }\hat{a}_{k}^{\dagger }+\frac{\Delta
_{j,k}^{\ast }}{2}\hat{a}_{j}\hat{a}_{k}),  \label{Has}
\end{equation}%
where $h_{j,k}=h_{k,j}^{\ast }$, $\Delta _{j,k}=\Delta _{k,j}$,\ and $\hat{a}%
_{k}$ is the bosonic annihilation operator satisfying the bosonic commutation
relations $\left[ a_{j},a_{k}\right] =0$ and $\left[ a_{j}^{\dagger },a_{k}%
\right] =\delta _{jk}$. The Hamiltonian is Hermitian and reduces to the
well-known tight-binding model when $\Delta =0$. The Heisenberg equation of
motion (HEOM) is 
\begin{equation}
\frac{d\hat{a}_{j}(t)}{dt}=i[\hat{H},\hat{a}_{j}(t)]=i\sum_{k}[-h_{j,k}\hat{a%
}_{k}(t)-\Delta _{j,k}\hat{a}_{k}^{\dagger }(t)],  \label{HE}
\end{equation}%
which can be rewritten as 
\begin{equation}
\qquad i\frac{d}{dt}\hat{V}(t)=H_{D}\hat{V}(t).
\end{equation}%
with 
\begin{equation}
H_{D}=\left( 
\begin{array}{cc}
h & \Delta \\ 
-\Delta ^{\ast } & -h^{\ast }%
\end{array}%
\right) ,
\end{equation}%
\begin{equation}
\hat{V}(t)=[\hat{a}_{1}(t),\hat{a}_{2}(t),...,\hat{a}_{N}(t),\hat{a}%
_{1}^{\dagger }(t),\hat{a}_{2}^{\dagger }(t),...,\hat{a}_{N}^{\dagger
}(t)]^{T},
\end{equation}%
$h$ and $\Delta $ being $N\times N$ matrices. $H_{D}$ is the \textit{%
dynamical matrix}, instead of the Hamiltonian matrix. $H_{D}$ satisfies $\tau
_{x}H_{D}\tau _{x}=-H_{D}^{\ast }$, $\tau _{z}H_{D}\tau _{z}=H_{D}^{\dagger
} $ and $\tau _{y}H_{D}\tau _{y}=-H_{D}^{\mathbf{T}}$, where $\tau _{x},\tau
_{y}$, and $\tau _{z}$ are Pauli matrices.

According to HEOM, we have 
\begin{equation}
\hat{V}(t)=e^{-iH_{D}t}\hat{V}(0)=S(t)\hat{V}(0),
\end{equation}%
where 
\begin{equation}
S(t)=e^{-iH_{D}t}.
\end{equation}%
Thus the time evolution of the field operators becomes 
\begin{equation}
\hat{a}_{j}(t)=\sum_{k=1}^{N}\left( P_{j,k}(t)\hat{a}_{k}(0)+Q_{j,k}(t)\hat{a%
}_{k}^{\dagger }(0)\right) ,  \label{at}
\end{equation}%
where $P_{j,k}(t)=S_{j,k}(t)$, $Q_{j,k}(t)=S_{j,(k+N)}(t)$, and $j=1,2,...,N$%
. For convenience, we use $\hat{a}_{k}$ and $\hat{a}_{k}^{\dagger }$ to
represent $\hat{a}_{k}(0)$ and $\hat{a}_{k}^{\dagger }(0)$, respectively.

In this section, we consider a coherent initial state 
\begin{equation*}
|\psi _{0}\rangle =|\alpha _{1},\alpha _{2},...,\alpha _{N}\rangle =\hat{D}%
_{1}(\alpha _{1})\hat{D}_{2}(\alpha _{2})...\hat{D}_{N}(\alpha
_{N})|0\rangle ,
\end{equation*}%
where 
\begin{equation*}
\hat{D}_{j}(\alpha _{j})=e^{\alpha _{j}\hat{a}_{j}^{\dagger }-\alpha
_{j}^{\ast }\hat{a}_{j}}.
\end{equation*}

The QFI is defined as%
\begin{equation*}
F_{\eta }\left( t\right) =4[\langle \partial _{\eta }\psi _{t}||\partial
_{\eta }\psi _{t}\rangle -|\langle \psi _{t}||\partial _{\eta }\psi
_{t}\rangle |^{2}],
\end{equation*}%
which requires the evaluation of the quantum state at time $t$ 
\begin{equation*}
|\psi _{t}\rangle =e^{-i\hat{H}t}|\psi _{0}\rangle .
\end{equation*}%
Inset the coherent initial state in, we have 
\begin{equation*}
|\partial _{\eta }\psi _{t}\rangle =\partial _{\eta }e^{-i\hat{H}t}\hat{D}%
_{1}(\alpha _{1})\hat{D}_{2}(\alpha _{2})...\hat{D}_{N}(\alpha
_{N})|0\rangle .
\end{equation*}%
The derivative with the sensing parameter $\eta $ is only applied to the
term $e^{-i\hat{H}t}$ with  
\begin{equation}
\begin{split}
& \partial _{\eta }e^{-i\hat{H}t} \\
=& \int_{x=0}^{1}dx\cdot e^{-i\hat{H}t(1-x)}\partial _{\eta }(-i\hat{H}%
t)e^{-i\hat{H}tx} \\
=& -it\int_{x=0}^{1}dx\cdot e^{-i\hat{H}t(1-x)}\partial _{\eta }\left[
\sum_{m,n=1}^{N}(h_{m,n}\hat{a}_{m}^{\dagger }\hat{a}_{n}+\frac{\Delta _{m,n}%
}{2}\hat{a}_{m}^{\dagger }\hat{a}_{n}^{\dagger }+\frac{\Delta _{m,n}^{\ast }%
}{2}\hat{a}_{m}\hat{a}_{n})\right] e^{-i\hat{H}tx} \\
=& \sum_{m,n=1}^{N}-ite^{-i\hat{H}t}\int_{x=0}^{1}dx\cdot e^{i\hat{H}%
tx}\left( \partial _{\eta }h_{m,n}\hat{a}_{m}^{\dagger }\hat{a}_{n}+\frac{%
\partial _{\eta }\Delta _{m,n}}{2}\hat{a}_{m}^{\dagger }\hat{a}_{n}^{\dagger
}+\frac{\partial _{\eta }\Delta _{m,n}^{\ast }}{2}\hat{a}_{m}\hat{a}%
_{n}\right) e^{-i\hat{H}tx} \\
=& \sum_{m,n=1}^{N}-ite^{-i\hat{H}t}\int_{x=0}^{1}dx\cdot \left[ \partial
_{\eta }h_{m,n}\hat{a}_{m}^{\dagger }(xt)\hat{a}_{n}(xt)+\frac{\partial
_{\eta }\Delta _{m,n}}{2}\hat{a}_{m}^{\dagger }(xt)\hat{a}_{n}^{\dagger
}(xt)+\frac{\partial _{\eta }\Delta _{m,n}^{\ast }}{2}\hat{a}_{m}(xt)\hat{a}%
_{n}(xt)\right]  \\
& =\sum_{m,n=1}^{N}-ie^{-i\hat{H}t}\int_{y=0}^{t}dy\cdot \left[ \partial
_{\eta }h_{m,n}\hat{a}_{m}^{\dagger }(y)\hat{a}_{n}(y)+\frac{\partial _{\eta
}\Delta _{m,n}}{2}\hat{a}_{m}^{\dagger }(y)\hat{a}_{n}^{\dagger }(y)+\frac{%
\partial _{\eta }\Delta _{m,n}^{\ast }}{2}\hat{a}_{m}(y)\hat{a}_{n}(y)\right]
,
\end{split}
\label{U1}
\end{equation}%
where $y=xt$.

Substitute Eq. (\ref{at}) into Eq. (\ref{U1}) (for convenience, we use $%
P_{j,k}$ and $Q_{j,k}$ to represent $P_{j,k}(y)$ and $Q_{j,k}(y)$,
respectively), we have 
\begin{equation}
\begin{split}
& \partial _{\eta }e^{-i\hat{H}t} \\
=& \sum_{m,n,j,k=1}^{N}-ie^{-i\hat{H}t}\int_{y=0}^{t}dy\cdot \left[ \partial
_{\eta }h_{m,n}\left( P_{m,j}^{\ast }\hat{a}_{j}^{\dagger }+Q_{m,j}^{\ast }%
\hat{a}_{j}\right) \left( P_{n,k}\hat{a}_{k}+Q_{n,k}\hat{a}_{k}^{\dagger
}\right) +\right.  \\
& \left. \frac{\partial _{\eta }\Delta _{m,n}}{2}\left( P_{m,j}^{\ast }\hat{a%
}_{j}^{\dagger }+Q_{m,j}^{\ast }\hat{a}_{j}\right) \left( P_{n,k}^{\ast }%
\hat{a}_{k}^{\dagger }+Q_{n,k}^{\ast }\hat{a}_{k}\right) +\frac{\partial
_{\eta }\Delta _{m,n}^{\ast }}{2}\left( P_{m,j}\hat{a}_{j}+Q_{m,j}\hat{a}%
_{j}^{\dagger }\right) \left( P_{n,k}\hat{a}_{k}+Q_{n,k}\hat{a}_{k}^{\dagger
}\right) \right]  \\
=& \sum_{m,n,j,k=1}^{N}-ie^{-i\hat{H}t}\int_{y=0}^{t}dy\cdot \left[ \left(
\partial _{\eta }h_{m,n}P_{m,j}^{\ast }P_{n,k}+\frac{\partial _{\eta }\Delta
_{m,n}}{2}P_{m,j}^{\ast }Q_{n,k}^{\ast }+\frac{\partial _{\eta }\Delta
_{m,n}^{\ast }}{2}Q_{m,j}P_{n,k}\right) \hat{a}_{j}^{\dagger }\hat{a}%
_{k}+\left( \partial _{\eta }h_{m,n}P_{m,j}^{\ast }Q_{n,k}\right. \right.  \\
& \left. \left. +\frac{\partial _{\eta }\Delta _{m,n}}{2}P_{m,j}^{\ast
}P_{n,k}^{\ast }+\frac{\partial _{\eta }\Delta _{m,n}^{\ast }}{2}%
Q_{m,j}Q_{n,k}\right) \hat{a}_{j}^{\dagger }\hat{a}_{k}^{\dagger }+\left(
\partial _{\eta }h_{m,n}Q_{m,j}^{\ast }P_{n,k}+\frac{\partial _{\eta }\Delta
_{m,n}}{2}Q_{m,j}^{\ast }Q_{n,k}^{\ast }+\frac{\partial _{\eta }\Delta
_{m,n}^{\ast }}{2}P_{m,j}P_{n,k}\right) \hat{a}_{j}\hat{a}_{k}\right.  \\
& \left. +\left( \partial _{\eta }h_{m,n}Q_{m,j}^{\ast }Q_{n,k}+\frac{%
\partial _{\eta }\Delta _{m,n}}{2}Q_{m,j}^{\ast }P_{n,k}^{\ast }+\frac{%
\partial _{\eta }\Delta _{m,n}^{\ast }}{2}P_{m,j}Q_{n,k}\right) \hat{a}_{j}%
\hat{a}_{k}^{\dagger }\right]  \\
&
\end{split}%
\end{equation}%
\begin{equation}
\begin{split}
=& \sum_{m,n,j,k=1}^{N}-ie^{-i\hat{H}t}\int_{y=0}^{t}dy\cdot \left[ \left(
\partial _{\eta }h_{m,n}(P_{m,j}^{\ast }P_{n,k}+Q_{m,k}^{\ast }Q_{n,j})+%
\frac{\partial _{\eta }\Delta _{m,n}}{2}(P_{m,j}^{\ast }Q_{n,k}^{\ast
}+Q_{m,k}^{\ast }P_{n,j}^{\ast })+\right. \right.  \\
& \left. \left. \frac{\partial _{\eta }\Delta _{m,n}^{\ast }}{2}%
(Q_{m,j}P_{n,k}+P_{m,k}Q_{n,j})\right) \hat{a}_{j}^{\dagger }\hat{a}%
_{k}+\left( \partial _{\eta }h_{m,n}P_{m,j}^{\ast }Q_{n,k}+\frac{\partial
_{\eta }\Delta _{m,n}}{2}P_{m,j}^{\ast }P_{n,k}^{\ast }+\frac{\partial
_{\eta }\Delta _{m,n}^{\ast }}{2}Q_{m,j}Q_{n,k}\right) \hat{a}_{j}^{\dagger }%
\hat{a}_{k}^{\dagger }+\right.  \\
& \left. \left( \partial _{\eta }h_{m,n}Q_{m,j}^{\ast }P_{n,k}+\frac{%
\partial _{\eta }\Delta _{m,n}}{2}Q_{m,j}^{\ast }Q_{n,k}^{\ast }+\frac{%
\partial _{\eta }\Delta _{m,n}^{\ast }}{2}P_{m,j}P_{n,k}\right) \hat{a}_{j}%
\hat{a}_{k}+\right.  \\
& \left. \left( \partial _{\eta }h_{m,n}Q_{m,j}^{\ast }Q_{n,k}+\frac{%
\partial _{\eta }\Delta _{m,n}}{2}Q_{m,j}^{\ast }P_{n,k}^{\ast }+\frac{%
\partial _{\eta }\Delta _{m,n}^{\ast }}{2}P_{m,j}Q_{n,k}\right) \delta _{j,k}%
\right]  \\
=& -\frac{i}{2}e^{-i\hat{H}t}\left[ C_{0}+\sum_{k,j}\left( 2C_{1,k,j}\hat{a}%
_{k}^{\dagger }\hat{a}_{j}+C_{2,k,j}\hat{a}_{k}^{\dagger }\hat{a}%
_{j}^{\dagger }+C_{3,k,j}\hat{a}_{k}\hat{a}_{j}\right) \right]  \\
=& e^{-i\hat{H}t}\hat{\mathcal{O}}.
\end{split}%
\end{equation}%
Here 
\begin{equation}
\begin{split}
C_{0}=& \sum_{m,n,j=1}^{N}2\int_{y=0}^{t}dy\left( \partial _{\eta
}h_{m,n}Q_{m,j}^{\ast }Q_{n,j}+\frac{\partial _{\eta }\Delta _{m,n}}{2}%
Q_{m,j}^{\ast }P_{n,j}^{\ast }+\frac{\partial _{\eta }\Delta _{m,n}^{\ast }}{%
2}P_{m,j}Q_{n,j}\right)  \\
C_{1,j,k}=& \sum_{m,n=1}^{N}\int_{y=0}^{t}dy\left[ \partial _{\eta
}h_{m,n}(P_{m,j}^{\ast }P_{n,k}+Q_{m,k}^{\ast }Q_{n,j})+\frac{\partial
_{\eta }\Delta _{m,n}}{2}(P_{m,j}^{\ast }Q_{n,k}^{\ast }+Q_{m,k}^{\ast
}P_{n,j}^{\ast })+\right.  \\
& \left. \frac{\partial _{\eta }\Delta _{m,n}^{\ast }}{2}%
(Q_{m,j}P_{n,k}+P_{m,k}Q_{n,j})\right]  \\
C_{2,j,k}=& \sum_{m,n=1}^{N}\int_{y=0}^{t}dy\left[ \partial _{\eta
}h_{m,n}\left( P_{m,j}^{\ast }Q_{n,k}+P_{m,k}^{\ast }Q_{n,j}\right) +\frac{%
\partial _{\eta }\Delta _{m,n}}{2}\left( P_{m,j}^{\ast }P_{n,k}^{\ast
}+P_{m,k}^{\ast }P_{n,j}^{\ast }\right) +\right.  \\
& \left. \frac{\partial _{\eta }\Delta _{m,n}^{\ast }}{2}\left(
Q_{m,j}Q_{n,k}+Q_{m,k}Q_{n,j}\right) \right]  \\
C_{3,j,k}=& \sum_{m,n=1}^{N}\int_{y=0}^{t}dy\left[ \partial _{\eta
}h_{m,n}\left( Q_{m,j}^{\ast }P_{n,k}+Q_{m,k}^{\ast }P_{n,j}\right) +\frac{%
\partial _{\eta }\Delta _{m,n}}{2}\left( Q_{m,j}^{\ast }Q_{n,k}^{\ast
}+Q_{m,k}^{\ast }Q_{n,j}^{\ast }\right) +\right.  \\
& \left. \frac{\partial _{\eta }\Delta _{m,n}^{\ast }}{2}\left(
P_{m,j}P_{n,k}+P_{m,k}P_{n,j}\right) \right]  \\
\hat{\mathcal{O}}=& -\frac{i}{2}\left[ C_{0}+\sum_{k,j}\left( 2C_{1,k,j}\hat{%
a}_{k}^{\dagger }\hat{a}_{j}+C_{2,k,j}\hat{a}_{k}^{\dagger }\hat{a}%
_{j}^{\dagger }+C_{3,k,j}\hat{a}_{k}\hat{a}_{j}\right) \right] .
\end{split}
\label{co}
\end{equation}%
$\bm{C}_{z}$ ($z=1,2,3$), $P$, and $Q$ are $N\times N$ matrices and $[\bm{C}%
_{z}]_{j_{1},j_{2}}=C_{z,j_{1},j_{2}}$. In the matrix form, we can rewrite the above equations as  
\begin{equation}
\left( 
\begin{array}{cc}
\bm{C}_{1} & \bm{C}_{2} \\ 
\bm{C}_{2}^{\ast } & \bm{C}_{1}^{\ast }%
\end{array}%
\right) =\int_{y=0}^{t}dy[S(y)]^{\dagger }\Sigma _{z}\partial _{\eta
}H_{D}S(y),  \label{com}
\end{equation}%
where $\Sigma _{z}=\sigma _{z}\otimes \mathbb{I}$, $S(y)=e^{-iyH_{D}}$, $%
\sigma _{z}$ is a Pauli matrix, and $\mathbb{I}$ is the $N\times N$ identity matrix. 

The QFI becomes 
\begin{equation}
F_{\eta }=4[\langle \psi _{0}|\hat{\mathcal{O}}^{\dagger }\hat{\mathcal{O}}%
|\psi _{0}\rangle -|\langle \psi _{0}|\hat{\mathcal{O}}|\psi _{0}\rangle
|^{2}]. 
\label{QFI}
\end{equation}%
Notice that we have 
\begin{equation*}
\hat{D}^{-1}(\alpha _{j})\hat{a}_{j}\hat{D}(\alpha _{j})=e^{-\alpha _{j}\hat{%
a}_{j}^{\dagger }+\alpha _{j}^{\ast }\hat{a}_{j}}\hat{a}_{j}e^{\alpha _{j}%
\hat{a}_{j}^{\dagger }-\alpha _{j}^{\ast }\hat{a}_{j}}=\hat{a}_{j}+\alpha
_{j},
\end{equation*}%
\begin{equation}
\begin{split}
\hat{a}_{j}|\psi _{0}\rangle =& \hat{a}_{j}\hat{D}_{1}(\alpha _{1})\hat{D}%
_{2}(\alpha _{2})...\hat{D}_{j}(\alpha _{j})...\hat{D}_{N}(\alpha
_{N})|0\rangle  \\
=& \hat{D}_{j}(\alpha _{j})\hat{D}_{j}(-\alpha _{j})\hat{a}_{j}\hat{D}%
_{1}(\alpha _{1})\hat{D}_{2}(\alpha _{2})...\hat{D}_{j}(\alpha _{j})...\hat{D%
}_{N}(\alpha _{N})|0\rangle  \\
=& \alpha _{j}|\psi _{0}\rangle ,
\end{split}
\label{aph}
\end{equation}%
and 
\begin{equation}
\langle \psi _{0}|\hat{a}_{j}^{\dagger }=\alpha _{j}^{\ast }\langle \psi
_{0}|. 
\label{phad}
\end{equation}
Take Eq. (\ref{co}) into Eq. (\ref{QFI}), and we expand the first term of
Eq. (\ref{QFI}) as a summation of normal ordered terms (NOTs). According to
Eqs. (\ref{aph}) and (\ref{phad}), the normal ordered quadratic terms (NOQTs)
of the first and second terms of Eq. (\ref{QFI}) cancel with each other. We
have 
\begin{equation}
\begin{split}
& \langle \psi _{0}|\hat{\mathcal{O}}^{\dagger }\hat{\mathcal{O}}|\psi
_{0}\rangle  \\
=& \frac{1}{4}\langle \psi _{0}|\left[ C_{0}+%
\sum_{j_{1},j_{2}=1}^{N}(2C_{1,j_{2},j_{1}}^{\ast }\hat{a}_{j_{1}}^{\dagger }%
\hat{a}_{j_{2}}+C_{2,j_{1},j_{2}}^{\ast }\hat{a}_{j_{1}}\hat{a}%
_{j_{2}}+C_{3,j_{1},j_{2}}^{\ast }\hat{a}_{j_{1}}^{\dagger }\hat{a}%
_{j_{2}}^{\dagger })\right] \times  \\
& \left[ C_{0}+\sum_{k_{1},k_{2}=1}^{N}(2C_{1,k_{1},k_{2}}\hat{a}%
_{k_{1}}^{\dagger }\hat{a}_{k_{2}}+C_{2,k_{1},k_{2}}\hat{a}_{k_{1}}^{\dagger
}\hat{a}_{k_{2}}^{\dagger }+C_{3,k_{1},k_{2}}\hat{a}_{k_{1}}\hat{a}_{k_{2}})%
\right] |\psi _{0}\rangle  \\
=& \frac{1}{4}\langle \psi _{0}|\sum_{j_{1},j_{2},k_{1},k_{2}=1}^{N}\left(
4C_{1,j_{2},j_{1}}^{\ast }C_{1,k_{1},k_{2}}\hat{a}_{j_{1}}^{\dagger }\hat{a}%
_{j_{2}}\hat{a}_{k_{1}}^{\dagger }\hat{a}_{k_{2}}+2C_{1,j_{2},j_{1}}^{\ast
}C_{2,k_{1},k_{2}}\hat{a}_{j_{1}}^{\dagger }\hat{a}_{j_{2}}\hat{a}%
_{k_{1}}^{\dagger }\hat{a}_{k_{2}}^{\dagger }+\right.  \\
& \left. 2C_{2,j_{1},j_{2}}^{\ast }C_{1,k_{1},k_{2}}\hat{a}_{j_{1}}\hat{a}%
_{j_{2}}\hat{a}_{k_{1}}^{\dagger }\hat{a}_{k_{2}}+C_{2,j_{1},j_{2}}^{\ast
}C_{2,k_{1},k_{2}}\hat{a}_{j_{1}}\hat{a}_{j_{2}}\hat{a}_{k_{1}}^{\dagger }%
\hat{a}_{k_{2}}^{\dagger }+NOQTs\right) |\psi _{0}\rangle ,
\end{split}%
\label{odo}
\end{equation}%
\begin{align}
& \hat{a}_{j_{1}}^{\dagger }\hat{a}_{j_{2}}\hat{a}_{k_{1}}^{\dagger }\hat{a}%
_{k_{2}}=\hat{a}_{j_{1}}^{\dagger }\hat{a}_{k_{1}}^{\dagger }\hat{a}_{j_{2}}%
\hat{a}_{k_{2}}+\delta _{k_{1},j_{2}}\hat{a}_{j_{1}}^{\dagger }\hat{a}%
_{k_{2}},  \label{t1} \\
& \hat{a}_{j_{1}}^{\dagger }\hat{a}_{j_{2}}\hat{a}_{k_{1}}^{\dagger }\hat{a}%
_{k_{2}}^{\dagger }=\hat{a}_{j_{1}}^{\dagger }\hat{a}_{k_{1}}^{\dagger }\hat{%
a}_{j_{2}}\hat{a}_{k_{2}}^{\dagger }+\delta _{j_{2},k_{1}}\hat{a}%
_{j_{1}}^{\dagger }\hat{a}_{k_{2}}^{\dagger }=\hat{a}_{j_{1}}^{\dagger }\hat{%
a}_{k_{1}}^{\dagger }\hat{a}_{k_{2}}^{\dagger }\hat{a}_{j_{2}}+\delta
_{j_{2},k_{2}}\hat{a}_{j_{1}}^{\dagger }\hat{a}_{k_{1}}^{\dagger }+\delta
_{j_{2},k_{1}}\hat{a}_{j_{1}}^{\dagger }\hat{a}_{k_{2}}^{\dagger },
\label{t2} \\
& \hat{a}_{j_{1}}\hat{a}_{j_{2}}\hat{a}_{k_{1}}^{\dagger }\hat{a}_{k_{2}}=%
\hat{a}_{j_{1}}\hat{a}_{k_{1}}^{\dagger }\hat{a}_{j_{2}}\hat{a}%
_{k_{2}}+\delta _{k_{1},j_{2}}\hat{a}_{j_{1}}\hat{a}_{k_{2}}=\hat{a}%
_{k_{1}}^{\dagger }\hat{a}_{j_{1}}\hat{a}_{j_{2}}\hat{a}_{k_{2}}+\delta
_{j_{1},k_{1}}\hat{a}_{j_{2}}\hat{a}_{k_{2}}+\delta _{k_{1},j_{2}}\hat{a}%
_{j_{1}}\hat{a}_{k_{2}},  \label{t3}
\end{align}%
and 
\begin{equation}
\begin{split}
& \hat{a}_{j_{1}}\hat{a}_{j_{2}}\hat{a}_{k_{1}}^{\dagger }\hat{a}%
_{k_{2}}^{\dagger } \\
=& \hat{a}_{j_{1}}\hat{a}_{k_{1}}^{\dagger }\hat{a}_{j_{2}}\hat{a}%
_{k_{2}}^{\dagger }+\delta _{j_{2},k_{1}}\hat{a}_{j_{1}}\hat{a}%
_{k_{2}}^{\dagger } \\
=& \hat{a}_{k_{1}}^{\dagger }\hat{a}_{j_{1}}\hat{a}_{j_{2}}\hat{a}%
_{k_{2}}^{\dagger }+\delta _{j_{1},k_{1}}\hat{a}_{j_{2}}\hat{a}%
_{k_{2}}^{\dagger }+\delta _{j_{2},k_{1}}\hat{a}_{j_{1}}\hat{a}%
_{k_{2}}^{\dagger } \\
=& \hat{a}_{k_{1}}^{\dagger }\hat{a}_{j_{1}}\hat{a}_{k_{2}}^{\dagger }\hat{a}%
_{j_{2}}+\delta _{k_{2},j_{2}}\hat{a}_{k_{1}}^{\dagger }\hat{a}%
_{j_{1}}+\delta _{j_{1},k_{1}}\hat{a}_{j_{2}}\hat{a}_{k_{2}}^{\dagger
}+\delta _{j_{2},k_{1}}\hat{a}_{j_{1}}\hat{a}_{k_{2}}^{\dagger } \\
=& \hat{a}_{k_{1}}^{\dagger }\hat{a}_{k_{2}}^{\dagger }\hat{a}_{j_{1}}\hat{a}%
_{j_{2}}+\delta _{j_{1},k_{2}}\hat{a}_{k_{1}}^{\dagger }\hat{a}%
_{j_{2}}+\delta _{k_{2},j_{2}}\hat{a}_{k_{1}}^{\dagger }\hat{a}%
_{j_{1}}+\delta _{j_{1},k_{1}}\hat{a}_{j_{2}}\hat{a}_{k_{2}}^{\dagger
}+\delta _{j_{2},k_{1}}\hat{a}_{j_{1}}\hat{a}_{k_{2}}^{\dagger } \\
=& \hat{a}_{k_{1}}^{\dagger }\hat{a}_{k_{2}}^{\dagger }\hat{a}_{j_{1}}\hat{a}%
_{j_{2}}+\delta _{j_{1},k_{2}}\hat{a}_{k_{1}}^{\dagger }\hat{a}%
_{j_{2}}+\delta _{k_{2},j_{2}}\hat{a}_{k_{1}}^{\dagger }\hat{a}%
_{j_{1}}+\delta _{j_{1},k_{1}}\hat{a}_{k_{2}}^{\dagger }\hat{a}%
_{j_{2}}+\delta _{j_{2},k_{1}}\hat{a}_{k_{2}}^{\dagger }\hat{a}%
_{j_{1}}+\delta _{j_{1},k_{1}}\delta _{j_{2},k_{2}}+\delta
_{j_{2},k_{1}}\delta _{j_{1},k_{2}}. \\
& \label{t4}
\end{split}%
\end{equation}

Take Eqs. (\ref{co}) and (\ref{odo})-(\ref{t4}) into Eq. (\ref{QFI}), we can
get that 
\begin{equation*}
\begin{split}
F_{\eta }=& 4[\langle \psi _{0}|\hat{\mathcal{O}}^{\dagger }\hat{\mathcal{O}}%
|\psi _{0}\rangle -|\langle \psi _{0}|\hat{\mathcal{O}}|\psi _{0}\rangle
|^{2}] \\
=& 4\left\{ \frac{1}{4}\langle \psi
_{0}|\sum_{j_{1},j_{2},k_{1},k_{2}=1}^{N}\left( 4C_{1,j_{2},j_{1}}^{\ast
}C_{1,k_{1},k_{2}}\delta _{k_{1},j_{2}}\hat{a}_{j_{1}}^{\dagger }\hat{a}%
_{k_{2}}+2C_{1,j_{2},j_{1}}^{\ast }C_{2,k_{1},k_{2}}(\delta _{j_{2},k_{2}}%
\hat{a}_{j_{1}}^{\dagger }\hat{a}_{k_{1}}^{\dagger }+\delta _{j_{2},k_{1}}%
\hat{a}_{j_{1}}^{\dagger }\hat{a}_{k_{2}}^{\dagger })+\right. \right.  \\
& \left. \left. 2C_{2,j_{1},j_{2}}^{\ast }C_{1,k_{1},k_{2}}(\delta
_{j_{1},k_{1}}\hat{a}_{j_{2}}\hat{a}_{k_{2}}+\delta _{k_{1},j_{2}}\hat{a}%
_{j_{1}}\hat{a}_{k_{2}})+C_{2,j_{1},j_{2}}^{\ast }C_{2,k_{1},k_{2}}(\delta
_{j_{1},k_{2}}\hat{a}_{k_{1}}^{\dagger }\hat{a}_{j_{2}}+\delta _{k_{2},j_{2}}%
\hat{a}_{k_{1}}^{\dagger }\hat{a}_{j_{1}}+\delta _{j_{1},k_{1}}\hat{a}%
_{k_{2}}^{\dagger }\hat{a}_{j_{2}}+\right. \right.  \\
& \left. \left. \delta _{j_{2},k_{1}}\hat{a}_{k_{2}}^{\dagger }\hat{a}%
_{j_{1}}+\delta _{j_{1},k_{1}}\delta _{j_{2},k_{2}}+\delta
_{j_{2},k_{1}}\delta _{j_{1},k_{2}})+NOQT\right) |\psi _{0}\rangle -|\langle
\psi _{0}|\hat{\mathcal{O}}|\psi _{0}\rangle |^{2}\right\}  \\
=& \sum_{j_{1},j_{2},k_{1},k_{2}=1}^{N}\left[ 4C_{1,j_{2},j_{1}}^{\ast
}C_{1,k_{1},k_{2}}\delta _{k_{1},j_{2}}\alpha _{j_{1}}^{\ast }\alpha
_{k_{2}}+2C_{1,j_{2},j_{1}}^{\ast }C_{2,k_{1},k_{2}}(\delta
_{j_{2},k_{2}}\alpha _{j_{1}}^{\ast }\alpha _{k_{1}}^{\ast }+\delta
_{j_{2},k_{1}}\alpha _{j_{1}}^{\ast }\alpha _{k_{2}}^{\ast })+\right.  \\
& \left. 2C_{2,j_{1},j_{2}}^{\ast }C_{1,k_{1},k_{2}}(\delta
_{j_{1},k_{1}}\alpha _{j_{2}}\alpha _{k_{2}}+\delta _{k_{1},j_{2}}\alpha
_{j_{1}}\alpha _{k_{2}})+C_{2,j_{1},j_{2}}^{\ast }C_{2,k_{1},k_{2}}(\delta
_{j_{1},k_{2}}\alpha _{k_{1}}^{\ast }\alpha _{j_{2}}+\delta
_{k_{2},j_{2}}\alpha _{k_{1}}^{\ast }\alpha _{j_{1}}+\delta
_{j_{1},k_{1}}\alpha _{k_{2}}^{\ast }\alpha _{j_{2}}+\right.  \\
& \left. \delta _{j_{2},k_{1}}\alpha _{k_{2}}^{\ast }\alpha _{j_{1}}+\delta
_{j_{1},k_{1}}\delta _{j_{2},k_{2}}+\delta _{j_{2},k_{1}}\delta
_{j_{1},k_{2}})\right]  \\
&
\end{split}%
\end{equation*}%
\begin{equation}
\begin{split}
=& \sum_{j_{1},j_{2},k_{1}}\left( 4C_{1,j_{2},j_{1}}^{\ast
}C_{1,j_{2},k_{1}}\alpha _{j_{1}}^{\ast }\alpha
_{k_{1}}+2C_{1,j_{2},j_{1}}^{\ast }C_{2,k_{1},j_{2}}\alpha _{j_{1}}^{\ast
}\alpha _{k_{1}}^{\ast }+2C_{1,j_{2},j_{1}}^{\ast }C_{2,j_{2},k_{1}}\alpha
_{j_{1}}^{\ast }\alpha _{k_{1}}^{\ast }+\right.  \\
& \left. 2C_{2,j_{1},j_{2}}^{\ast }C_{1,j_{1},k_{1}}\alpha _{j_{2}}\alpha
_{k_{1}}+2C_{2,j_{1},j_{2}}^{\ast }C_{1,j_{2},k_{1}}\alpha _{j_{1}}\alpha
_{k_{1}}+C_{2,j_{1},j_{2}}^{\ast }C_{2,k_{1},j_{1}}\alpha _{k_{1}}^{\ast
}\alpha _{j_{2}}+C_{2,j_{1},j_{2}}^{\ast }C_{2,k_{1},j_{2}}\alpha
_{k_{1}}^{\ast }\alpha _{j_{1}}+\right.  \\
& \left. C_{2,j_{1},j_{2}}^{\ast }C_{2,j_{1},k_{1}}\alpha _{k_{1}}^{\ast
}\alpha _{j_{2}}+C_{2,j_{1},j_{2}}^{\ast }C_{2,j_{2},k_{1}}\alpha
_{k_{1}}^{\ast }\alpha _{j_{1}}\right) +\sum_{j_{1},j_{2}}\left(
C_{2,j_{1},j_{2}}^{\ast }C_{2,j_{1},j_{2}}+C_{2,j_{1},j_{2}}^{\ast
}C_{2,j_{2},j_{1}}\right) . \\
&
\end{split}
\label{fgss}
\end{equation}%
Notice that $C_{z,j_{1},j_{2}}=C_{z,j_{2},j_{1}}$ for any $z=2,3$ and $%
j_{1},j_{2}=1,2,...,N$ according to Eq. (\ref{co}), then we have 
\begin{equation}
\begin{split}
F_{\eta }=& \left[ \sum_{j_{1},j_{2},k_{1}}\left( 4C_{1,j_{2},j_{1}}^{\ast
}C_{1,j_{2},k_{1}}\alpha _{j_{1}}^{\ast }\alpha
_{k_{1}}+4C_{1,j_{2},j_{1}}^{\ast }C_{2,k_{1},j_{2}}\alpha _{j_{1}}^{\ast
}\alpha _{k_{1}}^{\ast }+4C_{2,j_{1},j_{2}}^{\ast }C_{1,j_{1},k_{1}}\alpha
_{j_{2}}\alpha _{k_{1}}+\right. \right.  \\
& \left. \left. 2C_{2,j_{1},j_{2}}^{\ast }C_{2,k_{1},j_{1}}\alpha
_{k_{1}}^{\ast }\alpha _{j_{2}}+2C_{2,j_{1},j_{2}}^{\ast
}C_{2,j_{1},k_{1}}\alpha _{k_{1}}^{\ast }\alpha _{j_{2}}\right)
+\sum_{j_{1},j_{2}}2C_{2,j_{1},j_{2}}^{\ast }C_{2,j_{1},j_{2}}\right]  \\
=& \sum_{j_{1},j_{2}}\left[ \sum_{k_{1}}\left( 4C_{1,j_{2},j_{1}}^{\ast
}C_{1,j_{2},k_{1}}\alpha _{j_{1}}^{\ast }\alpha _{k_{1}}+8\mathrm{Re}%
[C_{1,j_{2},j_{1}}^{\ast }C_{2,k_{1},j_{2}}\alpha _{j_{1}}^{\ast }\alpha
_{k_{1}}^{\ast }]+4C_{2,j_{1},j_{2}}^{\ast }C_{2,j_{1},k_{1}}\alpha
_{k_{1}}^{\ast }\alpha _{j_{2}}\right) +2|C_{2,j_{1},j_{2}}|^{2}\right] .
\end{split}
\label{fgss1}
\end{equation}%
In term of matrices, we get 
\begin{equation}
\begin{split}
F_{\eta }=& 4\bm{\alpha}^{\dagger }\bm{C}_{1}^{\dagger }\bm{C}_{1}\bm{\alpha}%
+8\mathrm{Re}[\bm{\alpha}^{\dagger }\bm{C}_{2}\bm{C}_{1}^{\ast }\bm{\alpha}%
^{\ast }]+4\bm{\alpha}^{\mathrm{T}}\bm{C}_{2}^{\dagger }\bm{C}_{2}\bm{\alpha}%
^{\ast }+2\mathrm{Tr}[\bm{C}_{2}^{\dagger }\bm{C}_{2}] \\
=& 4\bm{B}^{\dagger }\bm{B}+2\mathrm{Tr}[\bm{C}_{2}^{\dagger }\bm{C}_{2}].
\end{split}
\label{QFIc}
\end{equation}%
where $\bm{B}=\bm{C}_{1}\bm{\alpha}+\bm{C}_{2}\bm{\alpha}^{\ast }$ and $%
\bm{\alpha}=[\alpha _{1},\alpha _{2},...,\alpha _{N}]^{\mathrm{T}}$ and $%
\mathrm{T}$ is transpose operator. $\bm{C}_{1}$ and $\bm{C}_{2}$ are $%
N\times N$ matrices given in Eq. (\ref{com}).

\subsection{B. QFI for general initial state and time-dependent Hamiltonian}

Here we consider a general initial state $|\psi _{0}\rangle
=\sum_{j=1}^{l}f_{j}|\psi _{j}\rangle $, where $\sum_{j=1}^{l}|f_{j}|^{2}=1$%
, $|\psi _{j}\rangle =|\alpha _{1}^{j},\alpha _{2}^{j},...,\alpha
_{N}^{j}\rangle $, $\bm{\alpha}^{j}=[\alpha _{1}^{j},\alpha
_{2}^{j},...,\alpha _{N}^{j}]^{\mathrm{T}}$, and $\bm{\alpha}^{j}\neq %
\bm{\alpha}^{k}$ for $j\neq k$. The time-dependent Hamiltonian 
\begin{equation}
\hat{H}(t)=\sum_{j,k=1}^{N}(h_{j,k}(t)\hat{a}_{j}^{\dagger }\hat{a}_{k}+%
\frac{\Delta _{j,k}(t)}{2}\hat{a}_{j}^{\dagger }\hat{a}_{k}^{\dagger }+\frac{%
\Delta _{j,k}^{\ast }(t)}{2}\hat{a}_{j}\hat{a}_{k}).  \label{Hast}
\end{equation}%
The QFI is 
\begin{equation}
\begin{split}
F_{\eta }=& 4[\langle \psi _{0}|\hat{\mathcal{O}}^{\dagger }\hat{\mathcal{O}}%
|\psi _{0}\rangle -|\langle \psi _{0}|\hat{\mathcal{O}}|\psi _{0}\rangle
|^{2}] \\
=& 4[(\sum_{j=1}^{l}f_{j}^{\ast }\langle \psi _{j}|)\hat{\mathcal{O}}%
^{\dagger }\hat{\mathcal{O}}(\sum_{j=1}^{l}f_{k}|\psi _{k}\rangle
)-|(\sum_{j=1}^{l}f_{j}^{\ast }\langle \psi _{j}|)\hat{\mathcal{O}}%
(\sum_{j=1}^{l}f_{k}|\psi _{k}\rangle )|^{2}] \\
=& 4[\sum_{j,k=1}^{l}f_{j}^{\ast }f_{k}\langle \psi _{j}|\hat{\mathcal{O}}%
^{\dagger }\hat{\mathcal{O}}|\psi _{k}\rangle -(\sum_{j,k=1}^{l}f_{j}^{\ast
}f_{k}\langle \psi _{j}|\hat{\mathcal{O}}|\psi _{k}\rangle
)(\sum_{m,n=1}^{l}f_{m}^{\ast }f_{n}\langle \psi _{m}|\hat{\mathcal{O}}%
^{\dagger }|\psi _{n}\rangle )] \\
=& 4[\sum_{j,k=1}^{l}f_{j}^{\ast }f_{k}\langle \psi _{j}|\hat{\mathcal{O}}%
^{\dagger }\hat{\mathcal{O}}|\psi _{k}\rangle
-\sum_{j,k,m,n=1}^{l}f_{j}^{\ast }f_{k}f_{m}^{\ast }f_{n}\langle \psi _{j}|%
\hat{\mathcal{O}}|\psi _{k}\rangle \langle \psi _{m}|\hat{\mathcal{O}}%
^{\dagger }|\psi _{n}\rangle ] \\
&
\end{split}%
\end{equation}%
\begin{equation*}
\begin{split}
& \langle \psi _{j}|\hat{\mathcal{O}}|\psi _{k}\rangle  \\
=& \langle \psi _{j}|\left[ C_{0}+\sum_{k_{1},k_{2}=1}^{N}(2C_{1,k_{1},k_{2}}%
\hat{a}_{k_{1}}^{\dagger }\hat{a}_{k_{2}}+C_{2,k_{1},k_{2}}\hat{a}%
_{k_{1}}^{\dagger }\hat{a}_{k_{2}}^{\dagger }+C_{3,k_{1},k_{2}}\hat{a}%
_{k_{1}}\hat{a}_{k_{2}})\right] |\psi _{k}\rangle  \\
=& exp\left( -\frac{|\bm{\alpha}^{j}|^{2}+|\bm{\alpha}^{k}|^{2}}{2}+%
\bm{\alpha}^{j\ast }\cdot \bm{\alpha}^{k}\right) \left[ C_{0}+%
\sum_{k_{1},k_{2}=1}^{N}(2C_{1,k_{1},k_{2}}\alpha _{k_{1}}^{j\dagger }\alpha
_{k_{2}}^{k}+C_{2,k_{1},k_{2}}\alpha _{k_{1}}^{j\dagger }\alpha
_{k_{2}}^{j\dagger }+C_{3,k_{1},k_{2}}\alpha _{k_{1}}^{k}\alpha _{k_{2}}^{k})%
\right]  \\
=& exp\left( -\frac{|\bm{\alpha}^{j}|^{2}+|\bm{\alpha}^{k}|^{2}}{2}+%
\bm{\alpha}^{j\ast }\cdot \bm{\alpha}^{k}\right) \left[ C_{0}+2\bm{\alpha}%
^{j\dagger }\bm{C}_{1}\bm{\alpha}^{k}+\bm{\alpha}^{j\dagger }\bm{C}_{2}%
\bm{\alpha}^{j\ast }+\bm{\alpha}^{k\mathrm{\mathbf{T}}}\bm{C}_{2}^{\ast }%
\bm{\alpha}^{k}\right] ,
\end{split}%
\end{equation*}%
where we use $\bm{C}_{2}^{\ast }=\bm{C}_{3}$. 
\begin{equation}
\begin{split}
& \langle \psi _{m}|\hat{\mathcal{O}}^{\dagger }|\psi _{n}\rangle  \\
=& \langle \psi _{n}|\hat{\mathcal{O}}|\psi _{m}\rangle ^{\dagger } \\
=& exp\left( -\frac{|\bm{\alpha}^{m}|^{2}+|\bm{\alpha}^{n}|^{2}}{2}+%
\bm{\alpha}^{n}\cdot \bm{\alpha}^{m\ast }\right) (C_{0}+2\bm{\alpha}%
^{n\dagger }\bm{C}_{1}\bm{\alpha}^{m}+\bm{\alpha}^{n\dagger }\bm{C}_{2}%
\bm{\alpha}^{n\ast }+\bm{\alpha}^{m\mathrm{\mathbf{T}}}\bm{C}_{3}\bm{\alpha}%
^{m})^{\dagger } \\
=& exp\left( -\frac{|\bm{\alpha}^{m}|^{2}+|\bm{\alpha}^{n}|^{2}}{2}+%
\bm{\alpha}^{n}\cdot \bm{\alpha}^{m\ast }\right) \left[ C_{0}+2\bm{\alpha}%
^{m\dagger }\bm{C}_{1}\bm{\alpha}^{n}+\bm{\alpha}^{n\mathrm{\mathbf{T}}}%
\bm{C}_{2}^{\ast }\bm{\alpha}^{n}+\bm{\alpha}^{m\dagger }\bm{C}_{2}%
\bm{\alpha}^{m\ast }\right] 
\end{split}%
\label{ojk}
\end{equation}

\begin{equation}
\begin{split}
& \langle \psi _{j}|\hat{\mathcal{O}}^{\dagger }\hat{\mathcal{O}}|\psi
_{k}\rangle  \\
=& \frac{1}{4}\langle \psi _{j}|\left[ C_{0}+%
\sum_{j_{1},j_{2}=1}^{N}(2C_{1,j_{2},j_{1}}^{\ast }\hat{a}_{j_{1}}^{\dagger }%
\hat{a}_{j_{2}}+C_{2,j_{1},j_{2}}^{\ast }\hat{a}_{j_{1}}\hat{a}%
_{j_{2}}+C_{3,j_{1},j_{2}}^{\ast }\hat{a}_{j_{1}}^{\dagger }\hat{a}%
_{j_{2}}^{\dagger })\right] \times  \\
& \left[ C_{0}+\sum_{k_{1},k_{2}=1}^{N}(2C_{1,k_{1},k_{2}}\hat{a}%
_{k_{1}}^{\dagger }\hat{a}_{k_{2}}+C_{2,k_{1},k_{2}}\hat{a}_{k_{1}}^{\dagger
}\hat{a}_{k_{2}}^{\dagger }+C_{3,k_{1},k_{2}}\hat{a}_{k_{1}}\hat{a}_{k_{2}})%
\right] |\psi _{k}\rangle  \\
=& \frac{1}{4}\langle \psi _{j}|\sum_{j_{1},j_{2},k_{1},k_{2}=1}^{N}\left(
4C_{1,j_{2},j_{1}}^{\ast }C_{1,k_{1},k_{2}}\hat{a}_{j_{1}}^{\dagger }\hat{a}%
_{j_{2}}\hat{a}_{k_{1}}^{\dagger }\hat{a}_{k_{2}}+2C_{1,j_{2},j_{1}}^{\ast
}C_{2,k_{1},k_{2}}\hat{a}_{j_{1}}^{\dagger }\hat{a}_{j_{2}}\hat{a}%
_{k_{1}}^{\dagger }\hat{a}_{k_{2}}^{\dagger }+\right.  \\
& \left. 2C_{2,j_{1},j_{2}}^{\ast }C_{1,k_{1},k_{2}}\hat{a}_{j_{1}}\hat{a}%
_{j_{2}}\hat{a}_{k_{1}}^{\dagger }\hat{a}_{k_{2}}+C_{2,j_{1},j_{2}}^{\ast
}C_{2,k_{1},k_{2}}\hat{a}_{j_{1}}\hat{a}_{j_{2}}\hat{a}_{k_{1}}^{\dagger }%
\hat{a}_{k_{2}}^{\dagger }+NOQTs\right) |\psi _{k}\rangle.
\end{split}
\label{odjk}
\end{equation}%

For $R_{j,k}=4(\langle \psi _{j}|\hat{\mathcal{O}}^{\dagger }\hat{\mathcal{O}%
}|\psi _{k}\rangle -\langle \psi _{j}|\hat{\mathcal{O}}^{\dagger }|\psi
_{k}\rangle \langle \psi _{j}|\hat{\mathcal{O}}|\psi _{k}\rangle /\langle
\psi _{j}|\psi _{k}\rangle )$, the NOQT of the first term cancels with that
of the second term. We have
\begin{equation*}
\begin{split}
R_{j,k}=& 4[\langle \psi _{j}|\hat{\mathcal{O}}^{\dagger }\hat{\mathcal{O}}%
|\psi _{k}\rangle -\langle \psi _{j}|\hat{\mathcal{O}}^{\dagger }|\psi
_{k}\rangle \langle \psi _{j}|\hat{\mathcal{O}}|\psi _{k}\rangle /\langle
\psi _{j}|\psi _{k}\rangle ] \\
=& 4\left\{ \frac{1}{4}\langle \psi
_{j}|\sum_{j_{1},j_{2},k_{1},k_{2}=1}^{N}\left( 4C_{1,j_{2},j_{1}}^{\ast
}C_{1,k_{1},k_{2}}\delta _{k_{1},j_{2}}\hat{a}_{j_{1}}^{\dagger }\hat{a}%
_{k_{2}}+2C_{1,j_{2},j_{1}}^{\ast }C_{2,k_{1},k_{2}}(\delta _{j_{2},k_{2}}%
\hat{a}_{j_{1}}^{\dagger }\hat{a}_{k_{1}}^{\dagger }+\delta _{j_{2},k_{1}}%
\hat{a}_{j_{1}}^{\dagger }\hat{a}_{k_{2}}^{\dagger })+\right. \right.  \\
& \left. \left. 2C_{2,j_{1},j_{2}}^{\ast }C_{1,k_{1},k_{2}}(\delta
_{j_{1},k_{1}}\hat{a}_{j_{2}}\hat{a}_{k_{2}}+\delta _{k_{1},j_{2}}\hat{a}%
_{j_{1}}\hat{a}_{k_{2}})+C_{2,j_{1},j_{2}}^{\ast }C_{2,k_{1},k_{2}}(\delta
_{j_{1},k_{2}}\hat{a}_{k_{1}}^{\dagger }\hat{a}_{j_{2}}+\delta _{k_{2},j_{2}}%
\hat{a}_{k_{1}}^{\dagger }\hat{a}_{j_{1}}+\delta _{j_{1},k_{1}}\hat{a}%
_{k_{2}}^{\dagger }\hat{a}_{j_{2}}+\right. \right.  \\
& \left. \left. \delta _{j_{2},k_{1}}\hat{a}_{k_{2}}^{\dagger }\hat{a}%
_{j_{1}}^{k}+\delta _{j_{1},k_{1}}\delta _{j_{2},k_{2}}+\delta
_{j_{2},k_{1}}\delta _{j_{1},k_{2}})+NOQT\right) |\psi _{k}\rangle -\langle
\psi _{j}|\hat{\mathcal{O}}^{\dagger }|\psi _{k}\rangle \langle \psi _{j}|%
\hat{\mathcal{O}}|\psi _{k}/\langle \psi _{j}|\psi _{k}\rangle \rangle
)\right\}  \\
=& \sum_{j_{1},j_{2},k_{1},k_{2}=1}^{N}\langle \psi _{j}|\psi _{k}\rangle 
\left[ 4C_{1,j_{2},j_{1}}^{\ast }C_{1,k_{1},k_{2}}\delta
_{k_{1},j_{2}}\alpha _{j_{1}}^{j\ast }\alpha
_{k_{2}}^{k}+2C_{1,j_{2},j_{1}}^{\ast }C_{2,k_{1},k_{2}}(\delta
_{j_{2},k_{2}}\alpha _{j_{1}}^{j\ast }\alpha _{k_{1}}^{j\ast }+\delta
_{j_{2},k_{1}}\alpha _{j_{1}}^{j\ast }\alpha _{k_{2}}^{j\ast })+\right.  \\
& \left. 2C_{2,j_{1},j_{2}}^{\ast }C_{1,k_{1},k_{2}}(\delta
_{j_{1},k_{1}}\alpha _{j_{2}}^{k}\alpha _{k_{2}}^{k}+\delta
_{k_{1},j_{2}}\alpha _{j_{1}}^{k}\alpha
_{k_{2}}^{k})+C_{2,j_{1},j_{2}}^{\ast }C_{2,k_{1},k_{2}}(\delta
_{j_{1},k_{2}}\alpha _{k_{1}}^{j\ast }\alpha _{j_{2}}^{k}+\delta
_{k_{2},j_{2}}\alpha _{k_{1}}^{j\ast }\alpha _{j_{1}}^{k}+\delta
_{j_{1},k_{1}}\alpha _{k_{2}}^{j\ast }\alpha _{j_{2}}^{k}+\right.  \\
& \left. \delta _{j_{2},k_{1}}\alpha _{k_{2}}^{j\ast }\alpha
_{j_{1}}^{k}+\delta _{j_{1},k_{1}}\delta _{j_{2},k_{2}}+\delta
_{j_{2},k_{1}}\delta _{j_{1},k_{2}})\right]  \\
&
\end{split}%
\end{equation*}%
\begin{equation}
\begin{split}
=& \sum_{j_{1},j_{2},k_{1}}\langle \psi _{j}|\psi _{k}\rangle \left(
4C_{1,j_{2},j_{1}}^{\ast }C_{1,j_{2},k_{1}}\alpha _{j_{1}}^{j\ast }\alpha
_{k_{1}}^{k}+2C_{1,j_{2},j_{1}}^{\ast }C_{2,k_{1},j_{2}}\alpha
_{j_{1}}^{j\ast }\alpha _{k_{1}}^{j\ast }+2C_{1,j_{2},j_{1}}^{\ast
}C_{2,j_{2},k_{1}}\alpha _{j_{1}}^{j\ast }\alpha _{k_{1}}^{j\ast }+\right. 
\\
& \left. 2C_{2,j_{1},j_{2}}^{\ast }C_{1,j_{1},k_{1}}\alpha
_{j_{2}}^{k}\alpha _{k_{1}}^{k}+2C_{2,j_{1},j_{2}}^{\ast
}C_{1,j_{2},k_{1}}\alpha _{j_{1}}^{k}\alpha
_{k_{1}}^{k}+C_{2,j_{1},j_{2}}^{\ast }C_{2,k_{1},j_{1}}\alpha
_{k_{1}}^{j\ast }\alpha _{j_{2}}^{k}+C_{2,j_{1},j_{2}}^{\ast
}C_{2,k_{1},j_{2}}\alpha _{k_{1}}^{j\ast }\alpha _{j_{1}}^{k}+\right.  \\
& \left. C_{2,j_{1},j_{2}}^{\ast }C_{2,j_{1},k_{1}}\alpha _{k_{1}}^{j\ast
}\alpha _{j_{2}}^{k}+C_{2,j_{1},j_{2}}^{\ast }C_{2,j_{2},k_{1}}\alpha
_{k_{1}}^{j\ast }\alpha _{j_{1}}^{k}\right) +\sum_{j_{1},j_{2}}\langle \psi
_{j}|\psi _{k}\rangle \left( C_{2,j_{1},j_{2}}^{\ast
}C_{2,j_{1},j_{2}}+C_{2,j_{1},j_{2}}^{\ast }C_{2,j_{2},j_{1}}\right)  \\
=& \langle \psi _{j}|\psi _{k}\rangle \left[ \sum_{j_{1},j_{2},k_{1}}\left(
4C_{1,j_{2},j_{1}}^{\ast }C_{1,j_{2},k_{1}}\alpha _{j_{1}}^{j\ast }\alpha
_{k_{1}}^{k}+4C_{1,j_{2},j_{1}}^{\ast }C_{2,k_{1},j_{2}}\alpha
_{j_{1}}^{j\ast }\alpha _{k_{1}}^{j\ast }+4C_{2,j_{1},j_{2}}^{\ast
}C_{1,j_{1},k_{1}}\alpha _{j_{2}}^{k}\alpha _{k_{1}}^{k}+\right. \right.  \\
& \left. \left. 2C_{2,j_{1},j_{2}}^{\ast }C_{2,k_{1},j_{1}}\alpha
_{k_{1}}^{j\ast }\alpha _{j_{2}}^{k}+2C_{2,j_{1},j_{2}}^{\ast
}C_{2,j_{1},k_{1}}\alpha _{k_{1}}^{j\ast }\alpha _{j_{2}}^{k}\right)
+\sum_{j_{1},j_{2}}2C_{2,j_{1},j_{2}}^{\ast }C_{2,j_{1},j_{2}}\right]  \\
=& \sum_{j_{1},j_{2}}\langle \psi _{j}|\psi _{k}\rangle \left[
\sum_{k_{1}}4\left( C_{1,j_{2},j_{1}}^{\ast }C_{1,j_{2},k_{1}}\alpha
_{j_{1}}^{j\ast }\alpha _{k_{1}}^{k}+C_{1,j_{2},j_{1}}^{\ast
}C_{2,k_{1},j_{2}}\alpha _{j_{1}}^{j\ast }\alpha _{k_{1}}^{j\ast
}+C_{2,j_{1},j_{2}}^{\ast }C_{1,j_{1},k_{1}}\alpha _{j_{2}}^{k}\alpha
_{k_{1}}^{k}+\right. \right.  \\
& \left. \left. C_{2,j_{1},j_{2}}^{j\ast }C_{2,j_{1},k_{1}}\alpha
_{k_{1}}^{j\ast }\alpha _{j_{2}}^{k}\right) +2|C_{2,j_{1},j_{2}}|^{2}\right] 
\\
=& \langle \psi _{j}|\psi _{k}\rangle \left[ 4(\bm{\alpha}^{j\dagger }\bm{C}%
_{1}^{\dagger }\bm{C}_{1}\bm{\alpha}^{k}+\bm{\alpha}^{j\dagger }\bm{C}_{2}%
\bm{C}_{1}^{\ast }\bm{\alpha}^{j\ast }+\bm{\alpha}^{k\mathrm{\mathbf{T}}}%
\bm{C}_{2}^{\ast }\bm{C}_{1}\bm{\alpha}^{k}+\bm{\alpha}^{k\mathrm{T}}\bm{C}%
_{2}^{\dagger }\bm{C}_{2}\bm{\alpha}^{j\ast })+2\mathrm{Tr}[\bm{C}%
_{2}^{\dagger }\bm{C}_{2}]\right] .
\end{split}%
\label{rjk}
\end{equation}
\begin{equation}
\begin{split}
F_{\eta}=&4[\sum_{j,k=1}^l f_j^*f_k\langle\psi_j|\hat{\mathcal{O}}^{\dagger}%
\hat{\mathcal{O}} |\psi_k \rangle-\sum_{j,k,m,n=1}^l
f_j^*f_kf_m^*f_n\langle\psi_j|\hat{\mathcal{O}} |\psi_k \rangle
\langle\psi_m|\hat{\mathcal{O}}^{\dagger} |\psi_n \rangle] \\
=&\sum_{j,k=1}^l f_j^*f_k(R_{j,k}+4\langle\psi_j|\hat{\mathcal{O}}%
^{\dagger}|\psi_k \rangle\langle\psi_j|\hat{\mathcal{O}}|\psi_k
\rangle/\langle\psi_j|\psi_k \rangle)-\sum_{j,k,m,n=1}^l 4f_j^*f_kf_m^*f_n
\langle\psi_j|\hat{\mathcal{O}} |\psi_k \rangle \langle\psi_m|\hat{\mathcal{O%
}}^{\dagger} |\psi_n \rangle,
\end{split}%
\end{equation}
where $\langle\psi_j|\hat{\mathcal{O}} |\psi_k \rangle, \langle\psi_m|\hat{%
\mathcal{O}}^{\dagger} |\psi_n \rangle$, $R_{j,k}$ are given by Eqs. (\ref%
{ojk}), (\ref{odjk}), and (\ref{rjk}). $\langle\psi_j|\psi_k
\rangle=exp\left(-\frac{|\bm{\alpha}^j|^2+| \bm{\alpha}^k|^2}{2}+\bm{\alpha}%
^{j *}\cdot \bm{\alpha}^k \right)$.

\begin{equation}
\begin{split}
F_{\eta }=& \sum_{j,k=1}^{l}f_{j}^{\ast }f_{k}\langle \psi _{j}|\psi
_{k}\rangle \left\{ (\left[ 4(\bm{\alpha}^{j\dagger }\bm{C}_{1}^{\dagger }%
\bm{C}_{1}\bm{\alpha}^{k}+\bm{\alpha}^{j\dagger }\bm{C}_{2}\bm{C}_{1}^{\ast }%
\bm{\alpha}^{j\ast }+\bm{\alpha}^{k\mathrm{\mathbf{T}}}\bm{C}_{2}^{\ast }%
\bm{C}_{1}\bm{\alpha}^{k}+\bm{\alpha}^{k\mathrm{T}}\bm{C}_{2}^{\dagger }%
\bm{C}_{2}\bm{\alpha}^{j\ast })+2\mathrm{Tr}[\bm{C}_{2}^{\dagger }\bm{C}_{2}]%
\right] +\right.  \\
& \left. 4\left( C_{0}+2\bm{\alpha}^{j\dagger }\bm{C}_{1}\bm{\alpha}^{k}+%
\bm{\alpha}^{j\dagger }\bm{C}_{2}\bm{\alpha}^{j\ast }+\bm{\alpha}^{k\mathrm{%
\mathbf{T}}}\bm{C}_{2}^{\ast }\bm{\alpha}^{k}\right) \left( C_{0}+2%
\bm{\alpha}^{j\dagger }\bm{C}_{1}\bm{\alpha}^{k}+\bm{\alpha}^{k\mathrm{%
\mathbf{T}}}\bm{C}_{2}^{\ast }\bm{\alpha}^{k}+\bm{\alpha}^{j\dagger }\bm{C}%
_{2}\bm{\alpha}^{j\ast }\right) \right\} - \\
& \sum_{j,k,m,n=1}^{l}4f_{j}^{\ast }f_{k}f_{m}^{\ast }f_{n}\langle \psi
_{j}|\psi _{k}\rangle \langle \psi _{m}|\psi _{n}\rangle \left( C_{0}+2%
\bm{\alpha}^{j\dagger }\bm{C}_{1}\bm{\alpha}^{k}+\bm{\alpha}^{j\dagger }%
\bm{C}_{2}\bm{\alpha}^{j\ast }+\bm{\alpha}^{k\mathrm{\mathbf{T}}}\bm{C}%
_{2}^{\ast }\bm{\alpha}^{k}\right) \times  \\
& \left( C_{0}+2\bm{\alpha}^{m\dagger }\bm{C}_{1}\bm{\alpha}^{n}+\bm{\alpha}%
^{n\mathrm{\mathbf{T}}}\bm{C}_{2}^{\ast }\bm{\alpha}^{n}+\bm{\alpha}%
^{m\dagger }\bm{C}_{2}\bm{\alpha}^{m\ast }\right) 
\end{split}%
\end{equation}

Finally, consider an integral initial state $|\psi _{0}\rangle =\int_{-\infty }^{\infty
}dxf_{x}|\psi _{x}\rangle $, where $\int_{-\infty }^{\infty }dx|f_{x}|^{2}=1$%
, $|\psi _{x}\rangle =|\alpha _{1}^{x},\alpha _{2}^{x},...,\alpha
_{N}^{x}\rangle $, $\bm{\alpha}^{x}=[\alpha _{1}^{x},\alpha
_{2}^{x},...,\alpha _{N}^{x}]^{\mathrm{T}}$, and $\bm{\alpha}^{x}\neq %
\bm{\alpha}^{z}$ for $x\neq z$. The QFI is

\begin{equation}
\begin{split}
F_{\eta }=& \int_{-\infty }^{\infty }\int_{-\infty }^{\infty
}dxdzf_{x}^{\ast }f_{z}\langle \psi _{x}|\psi _{z}\rangle \left\{ \left[ 4(%
\bm{\alpha}^{x\dagger }\bm{C}_{1}^{\dagger }\bm{C}_{1}\bm{\alpha}^{z}+%
\bm{\alpha}^{x\dagger }\bm{C}_{2}\bm{C}_{1}^{\ast }\bm{\alpha}^{x\ast }+%
\bm{\alpha}^{z\mathrm{\mathbf{T}}}\bm{C}_{2}^{\ast }\bm{C}_{1}\bm{\alpha}%
^{z}+\bm{\alpha}^{z\mathrm{T}}\bm{C}_{2}^{\dagger }\bm{C}_{2}\bm{\alpha}%
^{x\ast })+\right. \right.  \\
& \left. \left. 2\mathrm{Tr}[\bm{C}_{2}^{\dagger }\bm{C}_{2}]\right]
+4\left( C_{0}+2\bm{\alpha}^{x\dagger }\bm{C}_{1}\bm{\alpha}^{z}+\bm{\alpha}%
^{x\dagger }\bm{C}_{2}\bm{\alpha}^{x\ast }+\bm{\alpha}^{z\mathrm{\mathbf{T}}}%
\bm{C}_{2}^{\ast }\bm{\alpha}^{z}\right) \left( C_{0}+2\bm{\alpha}^{x\dagger
}\bm{C}_{1}\bm{\alpha}^{z}+\bm{\alpha}^{z\mathrm{\mathbf{T}}}\bm{C}%
_{2}^{\ast }\bm{\alpha}^{z}+\bm{\alpha}^{x\dagger }\bm{C}_{2}\bm{\alpha}%
^{x\ast }\right) \right\} - \\
& \int_{-\infty }^{\infty }\int_{-\infty }^{\infty }\int_{-\infty }^{\infty
}\int_{-\infty }^{\infty }dxdzdyds4f_{x}^{\ast }f_{z}f_{y}^{\ast
}f_{s}\langle \psi _{x}|\psi _{z}\rangle \langle \psi _{y}|\psi _{s}\rangle
\left( C_{0}+2\bm{\alpha}^{x\dagger }\bm{C}_{1}\bm{\alpha}^{z}+\bm{\alpha}%
^{x\dagger }\bm{C}_{2}\bm{\alpha}^{x\ast }+\bm{\alpha}^{z\mathrm{\mathbf{T}}}%
\bm{C}_{2}^{\ast }\bm{\alpha}^{z}\right) \times  \\
& \left( C_{0}+2\bm{\alpha}^{y\dagger }\bm{C}_{1}\bm{\alpha}^{s}+\bm{\alpha}%
^{s\mathrm{\mathbf{T}}}\bm{C}_{2}^{\ast }\bm{\alpha}^{s}+\bm{\alpha}%
^{y\dagger }\bm{C}_{2}\bm{\alpha}^{y\ast }\right) .
\end{split}%
\end{equation}

\section{SII.~QFI and exceptional point sensitivity of a single mode sensor}

Consider a single mode Hamiltonian 
\begin{equation}
\hat{H}_{1}=\delta \hat{a}^{\dagger }\hat{a}+i\frac{\kappa }{2}(\hat{a}%
^{\dagger 2}-\hat{a}^{2}),  \label{Hss}
\end{equation}%
with the HEOM 
\begin{equation}
\frac{d\hat{a}}{dt}=i[\hat{H}_{1},\hat{a}]=-i\delta a+\kappa a^{\dagger }.
\label{Hess}
\end{equation}%
In the matrix form, we have 
\begin{equation}
\qquad i\frac{d}{dt}\hat{V}=H_{D1}\hat{V},
\end{equation}%
where $\hat{V}=\left( \hat{a},\hat{a}^{\dagger }\right) ^{T}$, and $%
H_{D1}=\delta \tau _{z}+i\kappa \tau _{x}$ is the dynamical matrix. The EP
is at $|\kappa |=|\delta |$. 

$\hat{V}(t)=e^{-iH_{D1}t}\hat{V}(0)=S\hat{V}(0)$
with 
\begin{equation}
S=e^{-iH_{D1}t}=\left( 
\begin{array}{cc}
P(t) & Q(t) \\ 
Q(t) & P^{\ast }(t)%
\end{array}%
\right) ,  \label{M}
\end{equation}%
where $P(t)$ and $Q(t)$ are given by Table (\ref{I}). 
\begin{table}[th]
\caption{The values of $P(t)$ and $Q(t)$.}
\begin{center}
\begin{tabular}{c|c|c}
\hline
& $P(t)$ & $Q(t)$ \\ \hline
$|\kappa|>|\delta|$ & $\cosh(\lambda_0t)-\frac{i\delta}{\lambda_0}%
\sinh(\lambda_0 t)$ & $\frac{\kappa}{\lambda_0}\sinh(\lambda_0t)$ \\ \hline
$|\kappa|<|\delta|$ & $\cos(\lambda_0t)-\frac{i\delta}{\lambda_0}%
\sin(\lambda_0 t)$ & $\frac{\kappa}{\lambda_0}\sin(\lambda_0t)$ \\ \hline
$|\kappa|=|\delta|$ & $1-it\delta$ & $t\delta$ \\ \hline
\end{tabular}%
\end{center}
\label{I}
\end{table}

For an initial state $|\psi _{0}\rangle =|\alpha \rangle =e^{\alpha \hat{a}%
^{\dagger }-\alpha ^{\ast }\hat{a}}|0\rangle $, $|\psi _{t}\rangle =e^{-i%
\hat{H}_{1}t}|\psi _{0}\rangle $, we have 

\begin{equation}
\begin{split}
F_{\kappa }& =4[\langle \partial _{\kappa }\psi _{t}||\partial _{\kappa
}\psi _{t}\rangle -|\langle \psi _{t}||\partial _{\kappa }\psi _{t}\rangle
|^{2}] \\
& =4|C_{1}\alpha +C_{2}\alpha ^{\ast }|^{2}+2|C_{2}|^{2},
\end{split}
\label{QFIss}
\end{equation}%
where 
\begin{equation}
\begin{split}
C_{1}=& i\int_{y=0}^{t}dy\left( P^{\ast }Q^{\ast }-PQ\right)  \\
C_{2}=& i\int_{y=0}^{t}dy\left( P^{\ast 2}-Q^{2}\right) .
\end{split}
\label{coss}
\end{equation}%
Take Table (\ref{I}) into Eq. (\ref{coss}), we get the values of $C_{1}$ and 
$C_{2}$ in Table (\ref{II}). 
\begin{table}[th]
\caption{The values of $C_{1}$ and $C_{2}$.}
\label{II}
\begin{center}
\begin{tabular}{c|c|c}
\hline
& $C_{1}$ & $C_{2}$ \\ \hline
$|\kappa|>|\delta|$ & $-\frac{\delta\kappa t}{\lambda_0^2}\left[\frac{%
\sinh(2\lambda_0t)}{2\lambda_0t}-1\right]$ & $\frac{2\delta^2t-i\delta}{%
2\lambda_0^2}-\frac{\kappa^2}{2\lambda_0^3}\sinh(2\lambda_0t)+\frac{i\delta}{%
2\lambda_0^2}\cosh(2\lambda_0t)$ \\ \hline
$|\kappa|<|\delta|$ & $-\frac{\delta\kappa t}{\lambda_0^2}\left[1-\frac{%
\sin(2\lambda_0t)}{2\lambda_0t}\right]$ & $\frac{i\delta-2\kappa^2t}{%
2\lambda_0^2}+\frac{\delta^2}{2\lambda_0^3}\sin(2\lambda_0t)-\frac{i\delta}{%
2\lambda_0^2}\cos(2\lambda_0t)$ \\ \hline
$|\kappa|=|\delta|$ & $-\frac{2\delta^2t^3}{3}$ & $it-\frac{2it^3\delta^2}{3}%
+\delta^2t^2$ \\ \hline
\end{tabular}%
\end{center}
\end{table}
\begin{figure}[tbph]
\centerline{\includegraphics[width=2in]{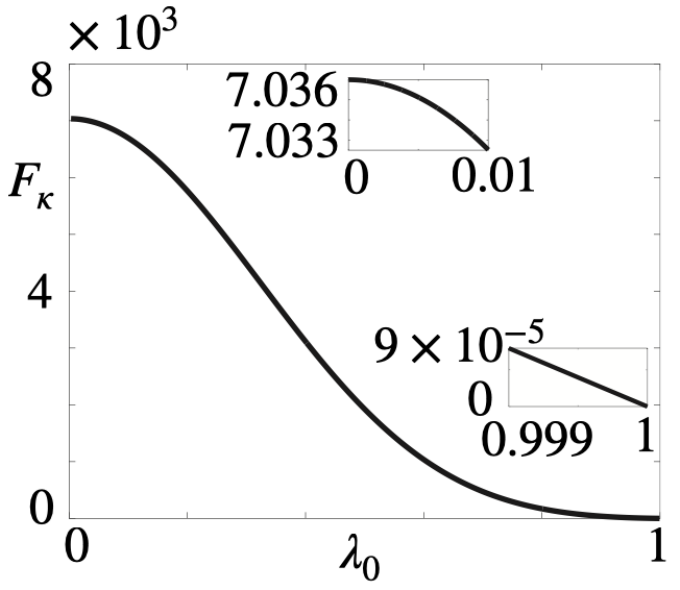}}
\caption{ QFI as a function of $\protect\lambda _{0}$ for different $\protect%
\lambda _{0}$ ranges. The parameters are $\protect\delta =1$ and $t=%
\protect\pi $. Top inset is a quadratic  function and bottom inset
is a linear function. }
\label{FS1}
\end{figure}

According to Eq. (\ref{QFIss}) and Table (\ref{II}), the QFI is finite
(without divergence) and continuous at or around the exceptional point (EP).
Figure \ref{FS1} shows QFI versus $\lambda _{0}$. We see QFI reaches it's
maximum at EP, and decreases to zero when $\lambda _{0}$ increases to $1$. $%
F_{\kappa }\propto t^{6}$ when $t\rightarrow +\infty $ at the EP. $F_{\kappa
}\propto t^{2}$ when $t\rightarrow +\infty $ for $|\delta |>|\kappa |$. $%
|\delta |<|\kappa |$ is an unstable region. The $t^{2}\rightarrow t^{6}$ time
scaling demonstrates that EP does enhance the sensitivity, but it is still
finite. 

At EP $|\kappa |=|\delta |$, The eigenvalues of $H_{D1}$ are 
\begin{equation}
\omega _{\pm }(0)=0.
\end{equation}%
Adding a perturbation to $\kappa $, i.e., $\kappa =\delta -\epsilon $ ($%
\epsilon \ll \delta $), the eigenvalues of $H_{D1}$ become $\omega _{\pm
}(\epsilon )$. The eigenvalue responses are $\Delta \omega _{\pm }=\omega
_{\pm }(\epsilon )-\omega _{\pm }(0)$. The leading orders of $\Delta \omega
_{\pm }$ are 
\begin{equation}
\Delta \omega _{+}=(2\kappa \epsilon )^{1/2},\quad \Delta \omega
_{-}=-(2\kappa \epsilon )^{1/2},  \label{De0}
\end{equation}%
which show a $\epsilon ^{1/2}$ scaling. The $\epsilon ^{1/2}$
scaling of spectrum response of the dynamical matrix corresponds
to the $t^{6}$ scaling of QFI. It satisfies $d_{F}=4M-2$ and $%
d_{\omega }=1/M$, where $M$ is the order of EP, $d_{F}$ is the scaling exponent
of the QFI (i.e., $F\propto t^{d_{F}}$ for $t\rightarrow \infty $) and $%
d_{\omega }$ is the scaling exponent of the maximum spectrum response of the
dynamical matrix [i.e., $\Delta \omega =\mathrm{max}_{\pm }(|\Delta \omega
_{\pm }|)\propto \epsilon ^{d_{\omega }}$ for $\epsilon \rightarrow 0$]. $%
d_{F}$ reaches the maximum response exponent predicted by the scaling law.

\section{SIII.~QFI and exceptional point sensitivity of a three mode sensor}

\subsection{A.~Without constraint}

Consider a three mode Hamiltonian 
\begin{equation}
\hat{H}_{3}=i\delta (\hat{a}_{1}^{\dagger }\hat{a}_{2}-\hat{a}_{2}^{\dagger }%
\hat{a}_{1}+\hat{a}_{2}^{\dagger }\hat{a}_{3}-\hat{a}_{3}^{\dagger }\hat{a}%
_{2})+\frac{i\kappa _{1}}{2}(\hat{a}_{1}^{2}-\hat{a}_{1}^{\dagger 2})+\frac{%
i\kappa _{3}}{2}(\hat{a}_{3}^{\dagger 2}-\hat{a}_{3}^{2}).  \label{Hts}
\end{equation}%
The HEOM is
\begin{equation}
\qquad i\frac{d}{dt}\hat{V}_{3}=H_{D3}\hat{V}_{3},
\end{equation}%
where $\hat{V}_{3}=\left( \hat{a}_{1},\hat{a}_{2},\hat{a}_{3},\hat{a}%
_{1}^{\dagger },\hat{a}_{2}^{\dagger },\hat{a}_{3}^{\dagger }\right) ^{T}$. $%
H_{D3}=\tau _{0}\otimes \mathbf{K}_{1}+\tau _{x}\otimes \mathbf{K}_{2}$ is
the dynamical matrix, where 
\begin{equation}
\mathbf{K}_{1}=i\left( 
\begin{array}{ccc}
0 & \delta  & 0 \\ 
-\delta  & 0 & \delta  \\ 
0 & -\delta  & 0%
\end{array}%
\right), \mathbf{K}_{2}=i\left( 
\begin{array}{ccc}
-\kappa _{1} & 0 & 0 \\ 
0 & 0 & 0 \\ 
0 & 0 & \kappa _{3}%
\end{array}%
\right) .
\end{equation}%
$\hat{V}_{3}(t)=e^{-iH_{D3}t}\hat{V}_{3}(0)=S\hat{V}_{3}(0)$, where $S=\tau
_{0}\otimes P(t)+\tau _{x}\otimes Q(t)$. 

Under a unitary transformation $(\tau _{x},\tau _{y},\tau _{z})\rightarrow
(\tau _{z},\tau _{y},-\tau _{x})$, $H_{D3}$ can be block diagonalized as $%
\tau _{0}\otimes \mathbf{K}_{1}+\tau _{z}\otimes \mathbf{K}_{2}$, with $%
H_{D3+}=\mathbf{K}_{1}+\mathbf{K}_{2}$ and $H_{D3-}=\mathbf{K}_{1}-\mathbf{K}%
_{2}$ being two irreducible blocks of $H_{D3}$. $H_{D3+}$ can be regarded as
the quantum generalization of the gain/neutral/loss cavity mode \cite%
{hodaei2017enhanced}. $H_{D3\pm }$ satisfies the parity time symmetry ($%
\mathcal{P}\mathcal{T}H_{D3\pm }\mathcal{T}^{-1}\mathcal{P}^{-1}=H_{D3\pm }$%
) given by 
\begin{equation}
\mathcal{P}=\left( 
\begin{array}{ccc}
0 & 0 & 1 \\ 
0 & 1 & 0 \\ 
1 & 0 & 0%
\end{array}%
\right) 
\end{equation}%
and $\mathcal{T}=\mathcal{K}$ is complex conjugate operator.

$H_{D3}$ possesses two degenerate third-order EP at $\sqrt{2}\delta =\kappa _{1}=\kappa
_{3}$, with the eigenvalues%
\begin{equation}
\omega _{j}^{\pm }(0)=0,\qquad \mathrm{where}\qquad j=1,2,3.
\end{equation}%
Adding a perturbation to $\kappa _{1}$, i.e., $\kappa _{1}=\sqrt{2}\delta
+\epsilon $ and $\kappa _{3}=\sqrt{2}\delta $ ($\epsilon \ll \delta $), the
spectrum responses are $\Delta \omega _{j}^{\pm }=\omega _{j}^{\pm
}(\epsilon )-\omega _{j}^{\pm }(0)$. Using Newton-Puiseux series expansion,
the leading orders of $\Delta \omega _{j}^{\pm }$ are 
\begin{equation}
\Delta \omega _{j}^{\pm }=ie^{i\pi j/3}\delta ^{2/3}\epsilon ^{1/3},
\end{equation}%
which have a $\epsilon ^{1/3}$ scaling.

Consider the initial state $|\psi _{0}\rangle =|\alpha _{1},\alpha
_{2},\alpha _{3}\rangle =D_{1}(\alpha _{1})D_{2}(\alpha _{2})D_{3}(\alpha
_{3})|0\rangle $ and $|\psi _{t}\rangle =e^{-i\hat{H}_{3}t}|\psi _{0}\rangle 
$. At the EP 
\begin{equation}
P(t)=\left( 
\begin{array}{ccc}
\frac{\delta ^{2}t^{2}}{2}+1 & \delta t & \frac{\delta ^{2}t^{2}}{2} \\ 
\delta (-t) & 1-\delta ^{2}t^{2} & \delta t \\ 
\frac{\delta ^{2}t^{2}}{2} & \delta (-t) & \frac{\delta ^{2}t^{2}}{2}+1 \\ 
\end{array}%
\right), Q(t)=\left( 
\begin{array}{ccc}
-\sqrt{2}\delta t & -\frac{\delta ^{2}t^{2}}{\sqrt{2}} & 0 \\ 
\frac{\delta ^{2}t^{2}}{\sqrt{2}} & 0 & \frac{\delta ^{2}t^{2}}{\sqrt{2}} \\ 
0 & -\frac{\delta ^{2}t^{2}}{\sqrt{2}} & \sqrt{2}\delta t \\ 
\end{array}%
\right) .  \label{pqe3}
\end{equation}

Take Eqs. (\ref{pqe3}) and (\ref{co}) into Eq. (\ref{QFIc}), we find $%
F_{\kappa _{1}}\propto t^{10}$ when $t\rightarrow +\infty $ at the
third-order EP. The $\epsilon ^{1/3}$ scaling of the spectrum response of
the dynamical matrix corresponds to $t^{10}$ scaling of the QFI. They also
satisfy $d_{F}=4/d_{\omega }-2$. 

Far from the EP at $\kappa _{1}=\kappa _{3}=0$, we have 
\begin{equation}
P(t)=\left( 
\begin{array}{ccc}
\cos ^{2}\left( \frac{\delta t}{\sqrt{2}}\right)  & \frac{\sin \left( \sqrt{2%
}\delta t\right) }{\sqrt{2}} & \sin ^{2}\left( \frac{\delta t}{\sqrt{2}}%
\right)  \\ 
-\frac{\sin \left( \sqrt{2}\delta t\right) }{\sqrt{2}} & \cos \left( \sqrt{2}%
\delta t\right)  & \frac{\sin \left( \sqrt{2}\delta t\right) }{\sqrt{2}} \\ 
\sin ^{2}\left( \frac{\delta t}{\sqrt{2}}\right)  & -\frac{\sin \left( \sqrt{%
2}\delta t\right) }{\sqrt{2}} & \cos ^{2}\left( \frac{\delta t}{\sqrt{2}}%
\right) 
\end{array}%
\right), Q(t)=\left( 
\begin{array}{ccc}
0 & 0 & 0 \\ 
0 & 0 & 0 \\ 
0 & 0 & 0%
\end{array}%
\right) .  \label{pqn3}
\end{equation}%
Take Eqs. (\ref{pqn3}) and (\ref{co}) into Eq. (\ref{QFIc}), we get that $%
F_{\kappa _{1}}\propto t^{2}$ when $t\rightarrow +\infty $ at $\kappa
_{1}=\kappa _{3}=0$. 

\subsection{B.~With constraint}

Consider the same three mode model in Eq. (\ref{Hts}) but with the
constraint $\kappa _{1}=\kappa _{2}=\eta $. The Hamiltonian becomes 
\begin{equation}
\hat{H}_{3}^{\prime }=i\delta (\hat{a}_{1}^{\dagger }\hat{a}_{2}-\hat{a}%
_{2}^{\dagger }\hat{a}_{1}+\hat{a}_{2}^{\dagger }\hat{a}_{3}-\hat{a}%
_{3}^{\dagger }\hat{a}_{2})+\frac{i\eta }{2}(\hat{a}_{1}^{2}-\hat{a}%
_{1}^{\dagger 2}+\hat{a}_{3}^{\dagger 2}-\hat{a}_{3}^{2}),  \label{Htcs}
\end{equation}%
with the HEOM
\begin{equation}
\qquad i\frac{d}{dt}\hat{V}_{3}=H_{D3}^{\prime }\hat{V}_{3},
\end{equation}%
where $\hat{V}_{3}=\left( \hat{a}_{1},\hat{a}_{2},\hat{a}_{3},\hat{a}%
_{1}^{\dagger },\hat{a}_{2}^{\dagger },\hat{a}_{3}^{\dagger }\right) ^{T}$.
The dynamical matrix $H_{D3}^{\prime }=\tau _{0}\otimes \mathbf{K}_{1}+\tau
_{x}\otimes \mathbf{K}_{2}$ with 
\begin{equation}
\mathbf{K}_{1}=i\left( 
\begin{array}{ccc}
0 & \delta  & 0 \\ 
-\delta  & 0 & \delta  \\ 
0 & -\delta  & 0%
\end{array}%
\right), \mathbf{K}_{2}=i\left( 
\begin{array}{ccc}
-\eta  & 0 & 0 \\ 
0 & 0 & 0 \\ 
0 & 0 & \eta 
\end{array}%
\right) .
\end{equation}%
$\hat{V}_{3}(t)=e^{-iH_{D3}^{\prime }t}\hat{V}_{3}(0)=S\hat{V}_{3}(0)$,
where $S=\tau _{0}\otimes P(t)+\tau _{x}\otimes Q(t)$. Add a perturbation to 
$\eta $, i.e., $\eta =\sqrt{2}\delta +\epsilon $ ($\epsilon \ll \delta $),
and use Newton-Puiseux series expansion, we find the leading orders of the
spectrum responses $\Delta \omega _{j}^{\pm }$ are 
\begin{equation}
\Delta \omega _{1}^{\pm }=0,\quad \Delta \omega _{2}^{\pm }=i\sqrt{2\sqrt{2}%
\delta \epsilon },\quad \Delta \omega _{3}^{\pm }=-i\sqrt{2\sqrt{2}\delta
\epsilon },
\end{equation}%
which show a $\epsilon ^{1/2}$ scaling for the maximum spectrum response. 

We consider the initial state $|\psi _{0}\rangle =|0\rangle $. $|\psi
_{t}\rangle =e^{-i\hat{H}_{3}^{\prime }t}|\psi _{0}\rangle $. At the
third-order EP $\sqrt{2}\delta =\eta $, we have Eq. (\ref{pqe3}). Take Eqs. (%
\ref{pqe3}) and (\ref{co}) into Eq. (\ref{QFIc}), we get that $F_{\eta
}\propto t^{6}$ when $t\rightarrow +\infty $ at the third-order EP. It
satisfies $d_{F}<4M-2$.

\section{SIV.~QFI scaling of bosonic Kitaev chain}

\subsection{A.~At exceptional point}

Consider a Hamiltonian $\hat{H}_{K}=\hat{H}_{BKC}+\hat{H}_{\eta }$,
where $\hat{H}_{BCK}$ is the bosonic Kitaev chain with 
\begin{equation}
\hat{H}_{BKC}=\sum_{j=1}^{N-1}\left( iJ\hat{a}_{j}^{\dagger }\hat{a}%
_{j+1}+i\Omega \hat{a}_{j}^{\dagger }\hat{a}_{j+1}^{\dagger }+h.c.\right) ,
\end{equation}%
and  
\begin{equation}
\hat{H}_{\eta }=i\eta \sin (\theta )\hat{a}_{N}^{\dagger }\hat{a}_{1}+i\eta
\cos (\theta )\hat{a}_{N}^{\dagger }\hat{a}_{1}^{\dagger }+h.c..
\end{equation}%
is the edge coupling term that couples modes 1 and $N$. The HEOM 
\begin{equation}
\qquad i\frac{d}{dt}\hat{V}_{K}=H_{DK}\hat{V}_{K},
\end{equation}%
with $\hat{V}_{K}=\left( \hat{a}_{1},\hat{a}_{2},...,\hat{a}_{N},\hat{a}%
_{1}^{\dagger },\hat{a}_{2}^{\dagger },...,\hat{a}_{N}^{\dagger }\right) ^{T}
$, and the dynamical matrix $H_{DK}=\tau _{0}\otimes \mathbf{L}_{1}+\tau
_{x}\otimes \mathbf{L}_{2}$. Here $\mathbf{L}_{1}$ and $\mathbf{L}_{2}$ are $%
N\times N$ matrices. $(\mathbf{L}_{1})_{N,1}=-(\mathbf{L}_{1})_{1,N}=i\eta
\sin (\theta )$, $(\mathbf{L}_{2})_{N,1}=(\mathbf{L}_{2})_{1,N}=i\eta \cos
(\theta )$, $(\mathbf{L}_{1})_{j,j+1}=-(\mathbf{L}_{1})_{j+1,j}=iJ$, and $(%
\mathbf{L}_{2})_{j,j+1}=(\mathbf{L}_{2})_{j+1,j}=i\Omega $, where $%
j=1,2,...,N-1$. $\theta \in \lbrack 0,2\pi )$. Other elements of $\mathbf{L}%
_{1}$ and $\mathbf{L}_{2}$ are $0$. Under unitary transformation $U=e^{-i%
\frac{\pi }{4}\tau _{y}}$, $(\tau _{x},\tau _{y},\tau _{z})\rightarrow (\tau
_{z},\tau _{y},-\tau _{x})$, and $H_{DK}$ can be block diagonalized as $\tau
_{0}\otimes \mathbf{L}_{1}+\tau _{z}\otimes \mathbf{L}_{2}$, with $H_{DK+}=%
\mathbf{L}_{1}+\mathbf{L}_{2}$ and $H_{DK-}=\mathbf{L}_{1}-\mathbf{L}_{2}$
being two irreducible blocks of $H_{DK}$. 

The eigenvalues of $H_{DK}$ at the $N$-th order EP $J=\Omega $, $\eta =0$ are 
\begin{equation}
\omega _{1}(0)=\omega _{2}(0)=...=\omega _{2N}(0)=0.
\end{equation}%
Add a perturbation to $\eta $, i.e., $\eta =\epsilon $ ($%
\epsilon \ll \Omega $), the eigenvalues of $H_{DK}$ are denoted as $\omega
_{1}(\epsilon )$, $\omega _{2}(\epsilon )$, ... $\omega _{2N}(\epsilon )$.
The spectrum response are $\Delta \omega _{j}=\omega _{j}(\epsilon
)-\omega _{j}(0)$, where $j=1,2,...,2N$. From the equation $H_{DK+}\psi
=E\Psi $, where $\Psi =[\psi _{1},\psi _{2},...,\psi _{N}]^{T}$, we find 
\begin{equation}
-iE\psi _{j}=2\Omega \psi _{j+1},for\quad j=2,3,...,N-1,
\label{e1}
\end{equation}%
\begin{equation}
-iE\psi _{1}=\epsilon \lbrack -\sin (\theta )+\cos (\theta )]\psi
_{N}+\epsilon \lbrack \sin (\theta )+\cos (\theta )]\psi _{2},
\label{e2}
\end{equation}%
\begin{equation}
-iE\psi _{N}=\epsilon \lbrack -\sin (\theta )+\cos (\theta )]\psi
_{N-1}+\epsilon \lbrack \sin (\theta )+\cos (\theta )]\psi _{1}.
\label{e3}
\end{equation}%
Combine Eqs. (\ref{e1})-(\ref{e3}) with $\psi _{1}=1$ (we can always have
that by fixing the gauge by dividing a constant for the eigenstates), we
have 
\begin{equation}
(2\Omega )^{2}z^{N}-\epsilon ^{2}\cos (2\theta )z^{N-2}-2\epsilon \Omega
\lbrack \cos (\theta )+\sin (\theta )]=0,\quad z=\frac{-iE}{2\Omega }.
\label{e4}
\end{equation}%
For $\theta \neq \frac{3\pi }{4}$ and $\frac{7\pi }{4}$, we can omit the
second term in Eq. (\ref{e4}), and the maximum spectrum response has a $\epsilon
^{1/N}$ scaling. Similar scaling also occurs for $H_{DK-}$.

Consider an initial vacuum state $|\psi _{0}\rangle =|0\rangle $ and $|\psi
_{t}\rangle =e^{-i\hat{H}_{K}t}|\psi _{0}\rangle $. At the $N$-th order EP $%
J=\Omega $ and $\eta =0$, we have 
\begin{equation*}
P(t)=\mathbb{I}+\sum_{p=1}^{N-1}\frac{(2\Omega t)^{p}(J_{+}^{p}+J_{-}^{p})}{%
2(p!)},Q(t)=\sum_{p=1}^{N-1}\frac{(2\Omega t)^{p}(J_{+}^{p}-J_{-}^{p})}{2(p!)%
},
\end{equation*}%
where $\mathbf{J}_{+}^{p}$ and $\mathbf{J}_{-}^{p}$ are $N\times N$
matrices. $(\mathbf{J}_{+}^{p})_{j,j+p}=1$, and $(\mathbf{J}%
_{-}^{p})_{j+p,j}=(-1)^{p}$, where $j=1,2,...,N-p$. Other elements of $%
\mathbf{J}_{+}^{p}$ and $\mathbf{J}_{-}^{p}$ are zero. Thus
\begin{equation}
\begin{split}
\lbrack P(t)]_{j_{1},j_{2}}=& (-1)^{(j_{1}-j_{2})\mathrm{H}(j_{1}-j_{2})}%
\frac{(2\Omega t)^{|j_{1}-j_{2}|}}{2[|j_{1}-j_{2}|!]}\quad for\quad
j_{1}\neq j_{2},\qquad \lbrack P(t)]_{j_{1},j_{1}}=1 \\
\lbrack Q(t)]_{j_{1},j_{2}}=& (-1)^{(j_{1}-j_{2})\mathrm{H}(j_{1}-j_{2})}%
\mathrm{sgn}(j_{2}-j_{1})\frac{(2\Omega t)^{|j_{1}-j_{2}|}}{2[|j_{1}-j_{2}|!]%
}\quad for\quad j_{1}\neq j_{2},\qquad \lbrack Q(t)]_{j_{1},j_{1}}=0.
\end{split}%
\label{pqen}
\end{equation}%
Here $\mathrm{H}(x)$ is Heaviside step function. Taking Eqs. (\ref{pqen}), (%
\ref{co}), and (\ref{com}) into Eq. (\ref{QFIc}), we get that $F_{\eta
}=\sum_{j,k=1}^{N}|C_{2,j,k}|^{2}$, where 
\begin{equation}
\begin{split}
C_{2,1,N}& =C_{2,N,1}=-\frac{i}{2}\left[ \cos (\theta )t+[\cos (\theta
)+\sin (\theta )]\frac{2(-1)^{N-1}(2\Omega )^{2(N-1)}}{(p!)^{2}}t^{2N-1}%
\right]  \\
C_{2,j,k}& =-i[\cos (\theta )+\sin (\theta )]\left[ \frac{(-1)^{N-j}(2\Omega
)^{N-j+k-1}t^{N-j+k}}{(N-j)!(k-1)!(N-j+k)}+\frac{(-1)^{N-k}(2\Omega
)^{N-k+j-1}t^{N-k+j}}{(N-k)!(j-1)!(N-k+j)}\right] , \\
C_{2,1,k}& =C_{2,k,1}=-\frac{i}{2}[\cos (\theta )+\sin (\theta )]\left[ 
\frac{2(-1)^{N-1}(2\Omega )^{N+k-2}t^{N+k-1}}{(N-1)!(k-1)!(N+k-1)}+\frac{%
(-1)^{N-k}(2\Omega )^{N-k}t^{N-k+1}}{(N-k)!(N-k+1)}\right] , \\
C_{2,N,k}& =C_{2,k,N}=-\frac{i}{2}[\cos (\theta )+\sin (\theta )]\left[ 
\frac{2(-1)^{N-k}(2\Omega )^{2N-k-1}t^{2N-k}}{(N-k)!(N-1)!(2N-k)}+\frac{%
(2\Omega )^{k-1}t^{k}}{(k-1)!k}\right] , \\
C_{2,1,1}& =-i[\cos (\theta )+\sin (\theta )]\frac{(2\Omega )^{N-1}t^{N}}{%
(N-1)!N} \\
C_{2,N,N}& =-i[\cos (\theta )+\sin (\theta )]\frac{(-2\Omega )^{N-1}t^{N}}{%
(N-1)!N},
\end{split}%
\end{equation}
$j,k=2,3,...,N-1$. When $t\rightarrow +\infty $ at the N-th order EP, $F_{\eta
}\propto t^{4N-2}$ for $\theta \neq \frac{3\pi }{4}$ and $\frac{7\pi }{4}$, and $%
F_{\eta }\propto t^{2}$ for $\theta =\frac{3\pi }{4}$ or $\frac{7\pi }{4}$. 
The scaling exponent satisfies $d_{F}=4M-2$ with $M=N$ here.

\subsection{B.~Region close to the exceptional point}

In this section, we derive the size ($N$) scaling of the QFI $F_{\eta }$ at $%
|J-\Omega |/|J+\Omega |\ll 1$ and $\eta =0$ region, which is close to the EP. Under the unitary matrix 
$U=e^{-i\frac{\pi }{4}\tau _{y}}$ for the block diagonalization of  $\hat{H}%
_{DK}$,
\begin{equation}
\begin{split}
U^{\dagger }H_{DK}U=& \tau _{0}\otimes \mathbf{L}_{1}+\tau _{z}\otimes 
\mathbf{L}_{2} \\
=& \mathrm{diag}[\mathbf{L}_{1}+\mathbf{L}_{2},\mathbf{L}_{1}-\mathbf{L}_{2}]
\\
=& \mathrm{diag}[-\tilde{J}S^{-1}\Sigma _{y}S,-\tilde{J}S\Sigma _{y}S^{-1}],
\end{split}%
\end{equation}%
where $\tilde{J}=\sqrt{(J-\Omega )(J+\Omega )}$, $S=diag[1,\beta ,\beta
^{2},...,\beta ^{N-1}]$, and $\beta =\sqrt{\frac{J+\Omega }{J-\Omega }}$. $%
\Sigma _{y}$ is a $N\times N$ matrix with $[\Sigma _{y}]_{j,j+1}=-[\Sigma
_{y}]_{j+1,j}=i$, and other elements  being $0$. Thus 
\begin{equation*}
e^{-iH_{DK}t}=U\mathrm{diag}[S^{-1}e^{i\tilde{J}\Sigma _{y}t}S,Se^{i\tilde{J}%
\Sigma _{y}t}S^{-1}]U^{\dagger }.
\end{equation*}%
Denote $\tilde{S}=e^{i\tilde{J}\Sigma _{y}t}$, then we have $[S^{-1}e^{i%
\tilde{J}\Sigma _{y}t}S]_{m,n}=\tilde{S}_{m,n}\beta ^{-m+n}$ , $[Se^{i\tilde{%
J}\Sigma _{y}t}S^{-1}]_{m,n}=\tilde{S}_{m,n}\beta ^{m-n}$,
and 
\begin{equation}
P_{m,n}=\tilde{S}_{m,n}\frac{\beta ^{-m+n}+\beta ^{m-n}}{2},\qquad \qquad
Q_{m,n}=\tilde{S}_{m,n}\frac{\beta ^{-m+n}-\beta ^{m-n}}{2}.
\label{pqne}
\end{equation}

Take Eqs. (\ref{pqne}), (\ref{co}), and (\ref{com}) into Eq. (\ref{QFIc}),
we find that the $\beta $'s leading order of the QFI is $\beta
^{4N-4}(\int_{0}^{t}dy|\tilde{S}_{N,1}|^{2})^{2}$ when $\int_{0}^{t}dy|%
\tilde{S}_{N,1}|^{2}\neq 0$. Therefore $F_{\eta }\propto \beta ^{4N-4}$.

\end{document}